\documentclass[12pt]{article} 

\usepackage[fleqn]{amsmath}
\usepackage{amsfonts}
\usepackage{amssymb}
\usepackage{mathrsfs}

\usepackage[dvipsnames]{xcolor}

\usepackage{epsfig}

\usepackage{caption}
\usepackage{float}
\usepackage[title]{appendix}

\usepackage{hyperref}
\hypersetup{
    colorlinks,
    citecolor=blue,
    filecolor=blue,
    linkcolor=blue,
    urlcolor=blue
}

\newcommand{\NN}{\mathbb{N}}
\newcommand{\RR}{\mathbb{R}}
\newcommand{\SP}{\mathbb{S}}
\newcommand{\ZZ}{\mathbb{Z}}

\newcommand{\beg}{\begin{equation}}
\newcommand{\en}{\end{equation}}

\newcommand{\eps}{\varepsilon}

\DeclareFontFamily{U}{mathx}{}
\DeclareFontShape{U}{mathx}{m}{n}{<-> mathx10}{}
\DeclareSymbolFont{mathx}{U}{mathx}{m}{n}
\DeclareMathAccent{\widecheck}{0}{mathx}{"71}

\begin{document}

\title{Bounds on $T_c$ in the Eliashberg theory of\\ Superconductivity. III: Einstein phonons\vspace{-0.5truecm}}
\author{M. K.-H. Kiessling,$^1$ B. L. Altshuler,$^2$  and E. A. Yuzbashyan$^3$\\ \small
          $^1$ Department of Mathematics,\\ \small
Rutgers, The State University of New Jersey,\\ \small
          110 Frelinghuysen Road, Piscataway, NJ 08854\\ \small
          $^2$ Department of Physics, \\ \small
Columbia University,\\ \small
             538 West 120th Street, New York, NY 10027\\ \small
             $^3$ Center of Materials Theory \\ \small Department of Physics and Astronomy, \\ \small
Rutgers, The State University of New Jersey,\\ \small
          136 Frelinghuysen Road, Piscataway, NJ 08854}

\date{Revised version of April 12, 2025}
\maketitle

\thispagestyle{empty}

\vfill\vfill
\hrule
\medskip
\noindent
\copyright(2025) 
\small{The authors. Reproduction of this preprint, in its entirety, is permitted for non-commercial purposes only.}

\newpage
\abstract{\noindent
 The dispersionless limit of the standard Eliashberg theory of superconductivity is studied,
in which the effective electron-electron interactions are mediated by 
Einstein phonons of frequency $\Omega>0$, equipped with electron-phonon coupling strength $\lambda$.
 The general results on $T_c$ for phonons with non-trivial dispersion relation,
obtained in a previous paper by the authors, (II), then become amenable to a detailed evaluation.
 The results are based on the traditional notion that the 
phase transition between normal and superconductivity coincides with 
the linear stability boundary $\mathscr{S}_{\!c}$ of the normal state region against 
perturbations toward the superconducting region.
 The variational principle for $\mathscr{S}_{\!c}$, obtained in (II), 
simplifies as follows: If $(\lambda,\Omega,T)\in\mathscr{S}_{\!c}$, then $\lambda = 1/\mathfrak{h}(\varpi)$, where 
$\varpi:=\Omega/2\pi T$, and where $\mathfrak{h}(\varpi)>0$ is the top eigenvalue of a compact self-adjoint 
operator $\mathfrak{H}(\varpi)$ on $\ell^2$ sequences;
$\mathfrak{H}(\varpi)$ is the dispersionless limit $P(d\omega)\to\delta(\omega-\Omega)d\omega$
of the operator $\mathfrak{K}(P,T)$ of (II).
 It is shown that when $\varpi \leq \sqrt{2}$, then the map $\varpi\mapsto\mathfrak{h}(\varpi)$ is invertible.
 For sufficiently large $\lambda$ ($\lambda>0.77$ will do) this yields the following:
(i) the existence of a critical temperature $T_c(\lambda,\Omega) = \Omega f(\lambda)$;
(ii) an ordered sequence of lower bounds on $f(\lambda)$ that converges to $f(\lambda)$.
 Also obtained is an upper bound on $T_c(\lambda,\Omega)$, which is not optimal yet agrees with the asymptotic 
behavior $T_c(\lambda,\Omega) \sim C \Omega\sqrt{\lambda}$ for large enough $\lambda$, given $\Omega$, 
though with a constant $C$ that is a factor $\approx 2.034$ larger than the optimal constant 
$\frac{1}{2\pi}\mathfrak{g}(2)^\frac12 =0.1827262477...$, with $\mathfrak{g}(\gamma)>0$ the largest eigenvalue 
of the compact self-adjoint operator $\mathfrak{G}(\gamma)$ for the $\gamma$ model, determined rigorously in the 
first one, (I), of this series of papers on $T_c$ by the authors.
}

\newpage

\section{Introduction}\label{sec:INTRO}\vspace{-.2truecm}

 This paper continues our rigorous inquiry into the critical temperature $T_c$ in the Eliashberg theory of superconductivity 
\cite{migdal,Eliashberg,BergmannRainer,AllenMitrovic,Carbotte,AllenDynes,Marsiglio}
that we initiated in \cite{KAYgamma}, where we also supplied a ``master introduction'' to this whole project, to which the 
reader is referred for general background information.
 In \cite{KAYgamma} we studied a version of this theory known as the $\gamma$ model, introduced recently by E.-G. Moon and
A. Chubukov \cite{MoonChubukov} (see also \cite{WAAYC}), which seeks to describe superconditivity in systems close to 
quantum phase transitions where the effective electron-electron interactions are 
mediated by collective bosonic excitations (fluctuations in the order parameter field).
 This effective interaction mechanism differs from the one in the standard version of Eliashberg theory where the 
effective electron-electron interactions are mediated by generally dispersive phonons of spectral density {(Eliashberg function)} $\alpha^2\!F(\omega)$ 
and electron-phonon coupling constant $\lambda:= 2\int_0^\infty\!\! \alpha^2\!F(\omega)\frac{d\omega}{\omega}$.
 Yet at $\gamma=2$ the $\gamma$ model captures the asymptotics at large coupling constant $\lambda$ of the
standard version of Eliashberg theory.
 In \cite{KAYdispersive} we studied $T_c$ in the standard version of Eliashberg theory,  building on our results obtained
in \cite{KAYgamma}.
 While the results obtained in \cite{KAYgamma} are quite explicit and quantitative, 
the results obtained in \cite{KAYdispersive} are rather qualitative, expressed in terms of integrals over the {Eliashberg} function 
$\alpha^2\!F(\omega)$ that was left largely unspecified except for some basic restrictions imposed by physical theory.
 To obtain more quantitative results within the standard version of Eliashberg theory, a detailed
specification of $\alpha^2\!F(\omega)$ is required.

  In the present paper we choose such a specification of $\alpha^2\!F(\omega)$ by considering the important dispersionless limit,
in which {$\alpha^2\!F(\omega)   \to \frac{\lambda\Omega}{2}\delta(\omega-\Omega) $}, 
featuring optical (Einstein) phonons of a single frequency $\Omega>0$.
 All the integrals over $\alpha^2\!F(\omega)$ in the results of \cite{KAYdispersive} then reduce to their integrands evaluated
at $\omega=\Omega$.
 This allows for much more detailed insights into the Eliashberg theory than would be possible with numerical quadratures 
of more than a half dozen {temperature}-dependent integrals over some spread-out function $\alpha^2\!F(\omega)$; of course, these more 
detailed insights are limited to the case of Einstein phonons and its immediate vicinity in the ``space of dispersion relations.''
 
 Incidentally, the dispersionless limit of the standard Eliashberg model is sometimes called the Holstein model, after
\cite{Holstein1}, \cite{Holstein2}; note though that in the Holstein model the bare phonons are dispersionless, while in 
the  Eliashberg model with Einstein phonons the renormalized phonons are.

 The Eliashberg model with Einstein phonons comes equipped with three parameters:
$\lambda>0$ and $\Omega>0$ are temperature-independent {material characteristics, while $T>0$ is the thermodynamic temperature.}
 After many years of (nonrigorous) theoretical and numerical work a ``thermodynamic narrative'' for the Eliashberg theory 
has emerged  \cite{AllenMitrovic,Carbotte,AllenDynes} that, for the version with Einstein phonons, can be summarized thus:
\smallskip

\noindent
{\bf Narrative}: \textsl{There is a critical temperature $T_c >0$, depending on $\lambda > 0$ and $\Omega >0$, 
such that for temperatures  $T\geq T_c$, the normal state is the unique thermal equilibrium phase
whereas at temperatures $T\in(0,T_c)$ a {\it{superconducting state}} is the unique thermal equilibrium phase,
up to an irrelevant gauge transformation.
 Moreover, the phase transition at $T_c$ from normal to superconductivity is continuous.} 
 \smallskip

 In our previous papers \cite{KAYgamma} and \cite{KAYdispersive} we took some steps toward the rigorous vindication of the 
analogous thermodynamic narrative for the $\gamma$ model and for the standard Eliashberg model with dispersive phonon model, 
with $\Omega>0$ replaced by $\gamma>0$, respectively by $P(d\omega)\in \mathcal{P}$, where $\mathcal{P}$ is the set of
(formal) probability measures over the positive frequencies $\omega\in \RR_+$ that have a density w.r.t. Lebesgue measure that
is $\propto \omega$ for small $\omega$ and vanishes for $\omega>\overline\Omega(P)$. 
 In the limit $P(d\omega)\to \delta(\omega-\Omega)d\omega$, with $\Omega>0$, our results in \cite{KAYdispersive} yield
the analogous (partial) vindication of the thermodynamical narrative stated above for the Eliashberg model with Einstein phonons.

 By ``partial vindication'' we primarily mean the following. 
 As in \cite{KAYdispersive}, we here \textsl{assume} the existence of a continuous phase transition between normal and 
superconductivity, so that its location in the phase diagram coincides with the linear-stability boundary $\mathscr{S}_{\!c}$ 
of the normal state region against perturbations toward the superconducting region.
 Thus the results of \cite{KAYgamma} and \cite{KAYdispersive}, and also those of the present paper that are obtained by 
specialization, are based on a rigorous study of the Eliashberg gap equations linearized about the normal state.
 As emphasized in \cite{KAYdispersive}, and already in \cite{KAYgamma}, 
a proper confirmation of the existence of a continuous transition between the normal and superconductivity phases requires
a study of the nonlinear Eliashberg gap equations, which we hope to present in a later publication.

 Another element of ``partial vindication'' is specific to \cite{KAYdispersive}, where the proof of existence of $T_c(\lambda,P)$
is restricted to $\lambda\geq \lambda_*(P)$, with $\lambda_*(P)$ given explicitly as an elementary function, though
involving more than a half dozen averages over $P(d\omega)$, left largely unspecified.
 For the Eliashberg model with Einstein phonons we will inherit this restriction to $\lambda\geq \lambda_*(P)$,
yet with $P(d\omega)=\delta(\omega-\Omega)d\omega$ the averages can be carried out explicitly.
 No analogous restriction occurs for the $\gamma$ model, where we proved the existence of $T_c(g,\gamma)$ for all $\gamma>0$, 
and any coupling constant $g>0$ of the $\gamma$ model.
 The restrictions on $\lambda$ expressed above are due to the technical limitations of our techniques of proof and not 
expected to be of any model-intrinsic significance.

 We next state our main results in more detail.
\vspace{-0.5truecm}

\section{The main results}

 Although most, though not all of our results in the present paper 
are special cases of the results we proved in \cite{KAYdispersive}, we state them as
theorems or propositions in their own right, rather than as corollaries. 

 The phase diagram we will be discussing in this paper 
consists of normal and superconducting thermal equilibrium regions in the positive $(\lambda,\Omega,T)$-octant.
 The results in \cite{KAYdispersive} yield the following theorem about these two regions.
\smallskip

\noindent
{\bf Theorem~1}: 
\textsl{The positive $(\lambda,\Omega,T)$-octant of the model consists of two simply connected regions.
 In one region the normal state is unstable against small perturbations toward the superconducting region, 
in the other region it is linearly stable.
 The boundary between the two regions, called the critical surface $\mathscr{S}_{\!c}$, is a graph over the positive
$(\Omega,T)$-quadrant, i.e.} $\mathscr{S}_{\!c}=\{(\lambda,\Omega,T)\in\RR_+^3: \lambda = \Lambda_{\mbox{\tiny{E}}}(\Omega,T)\}$. 
\textsl{The function} $\Lambda_{\mbox{\tiny{E}}}$ \textsl{is continuous and depends on $\Omega$ and $T$ only through the combination 
$\varpi:= \frac{\Omega}{2\pi T}$; thus,} $\Lambda_{\mbox{\tiny{E}}}(\Omega,T)=L_{\mbox{\tiny{E}}}(\varpi)$.
\textsl{The thermal equilibrium state at temperature $T$ of a 
crystal with Einstein phonon frequency $\Omega$ and electron-phonon coupling constant $\lambda$ is
the superconducting phase when} 
$\lambda>L_{\mbox{\tiny{E}}}(\varpi)$ \textsl{and the normal phase when} $\lambda<L_{\mbox{\tiny{E}}}(\varpi)$.
\smallskip

 Moreover, it follows from the results in \cite{KAYdispersive} that the function
$L_{\mbox{\tiny{E}}}(\varpi)$ is explicitly characterized by a variational principle.

\smallskip
\noindent
{\bf Theorem~2}: 
\textit{The function} $L_{\mbox{\tiny{E}}}(\varpi)$ \textsl{is determined by the following variational principle,}
\begin{equation}\label{eq:lambdaVP}
 L_{\mbox{\tiny{E}}}(\varpi) = \frac{1}{\mathfrak{h}(\varpi)}, 
\end{equation}
\textsl{where ${\mathfrak{h}(\varpi)}>0$ is the largest eigenvalue of an explicitly constructed 
compact self-adjoint operator $\mathfrak{H}(\varpi)$ on the Hilbert space of square-summable
sequences over the non-negative integers.}
\smallskip

 Our variational principle (\ref{eq:lambdaVP}) is obtained in the limit $P(d\omega)=\delta(\omega-\Omega)d\omega$ from the variational principle
$\lambda = 1/\mathfrak{k}(P,T)$, where $\mathfrak{k}(P,T)>0$ is the largest eigenvalue of a compact self-adjoint operator $\mathfrak{K}(P,T)$ 
constucted in \cite{KAYdispersive}.
 
 In \cite{KAYdispersive} we also discussed the approximation of 
$\mathfrak{K}(P,T)$ with a nested sequence of finite-rank operators that converges to $\mathfrak{K}(P,T)$,
and so obtained an increasing sequence of rigorous lower bounds on $\mathfrak{k}(P,T)$.
 The first four of these we computed in closed form, though involving up to seven $T$-dependent quadratures over 
$P(d\omega)$ that cannot be carried out without specification of $P$, and even then would in general require a numerical 
quadrature scheme.
 In the dispersionless limit $P(d\omega)=\delta(\omega-\Omega)d\omega$, these quadratures become trivial.
 This gives the following theorem.
\smallskip

\noindent
{\bf Theorem~3}: \textsl{For all} $N\in\NN$, $L_{\mbox{\tiny{E}}}(\varpi) < 1/\mathfrak{h}^{(N)}(\varpi)$, 
\textsl{where $\mathfrak{h}^{(N)}(\varpi)$ is the largest eigenvalue of $\mathfrak{H}^{(N)}(\varpi)$, the 
restriction of $\mathfrak{H}(\varpi)$ to the first $N$ components of $\ell^2({\NN}_0)$.
 The eigenvalues $\mathfrak{h}^{(N)}(\varpi)$ can be explicitly computed for $N\in\{1,2,3,4\}$.
 They read}
\begin{equation}\label{eq:hONE}
\mathfrak{h}^{(1)}(\varpi)= \frac{\varpi^2}{1+\varpi^2},
\end{equation}
\textsl{which is the one and only eigenvalue of $\mathfrak{H}^{(1)}$;}
\begin{equation}\label{eq:hTWO}
\mathfrak{h}^{(2)}(\varpi) =
\tfrac12\Big({\rm tr}\,\mathfrak{H}^{(2)} + \sqrt{\big({\rm tr}\,\mathfrak{H}^{(2)}\big)^2-4 \det\mathfrak{H}^{(2)}}\,\Big)(\varpi),
\end{equation}
\textsl{where $\mathfrak{H}^{(2)}(\varpi)$ is the upper leftmost $2\times2$ block of the matrix} $\mathfrak{H}^{(4)}(\varpi)$
\textsl{displayed further below;}
\begin{align}\label{eq:hTHREE}
\hspace{-0.7truecm}
\mathfrak{h}^{(3)}(\varpi)  = 
\tfrac13\left(\! \textstyle{ {\rm tr}\,\mathfrak{H}^{(3)} +
6\sqrt{\frac{p}{3}}\cos \left[\frac13\arccos\left(\frac{q}{2}\sqrt{\!\Big(\frac3p\Big)^{\!{}_3}}\,\right)\!\right] 
} \right)\!\!(\varpi)
,\!\!
\end{align}
\textsl{with  (temporarily suspending displaying the dependence on $\varpi$)}
\begin{align}\label{eq:3x3hP}
p = \tfrac13\big({\rm tr}\,\mathfrak{H}^{(3)}\big)^2- {\rm tr\, adj}\, \mathfrak{H}^{(3)},
\end{align}
\begin{align}\label{eq:3x3hQ}
q = 
 \tfrac{2}{27}\big({\rm tr}\,\mathfrak{H}^{(3)}\big)^3 
- \tfrac13 \big({\rm tr}\,\mathfrak{H}^{(3)}\big) \big({\rm tr\, adj}\,\mathfrak{H}^{(3)}\big)
+ \det\mathfrak{H}^{(3)},
\end{align}
\textsl{where $\mathfrak{H}^{(3)}(\varpi)$ is the upper leftmost $3\times3$ block of the matrix $\mathfrak{H}^{(4)}(\varpi)$
displayed further below;}
\begin{align}\label{eq:hFOUR}
\hspace{-0.8truecm}
\mathfrak{h}^{(4)}(\varpi)  = 
\Big[\! \sqrt{\tfrac12 Z}
+\!\sqrt{\!\tfrac{3}{16} A^2 -\tfrac12 B - \tfrac12 Z +\tfrac{A^3 -4AB + 8C}{16\sqrt{2Z}}}\!-\tfrac14 A \Big]\!(\varpi),
\end{align}
\textsl{where  $Z(\varpi)$ is a positive zero of the so-called \textsl{resolvent cubic} associated with the characteristic polynomial 
$\det \big(\eta\mathfrak{I} - \mathfrak{H}^{(4)}(\varpi)\big)$,
given by (temporarily suspending displaying the dependence on $\varpi$ again)}
\begin{equation}
\label{eq:Zdef}
Z = \tfrac13 \Big[ \sqrt{Y} \cos\Big(\tfrac13 \arccos \tfrac{X}{2\sqrt{Y^3}}\Big) - B + \tfrac38 A^2\Big],
\end{equation}
\textsl{with}
\begin{align}
\label{eq:Xdef}
X =&  2B^3 - 9ABC+27C^2 + 27A^2D -72BD,\\
\label{eq:Ydef}
Y =& B^2-3AC+12D,
\end{align}
\textsl{where}
\begin{align}
\label{eq:Adef}
A &= - {\rm tr}\, \mathfrak{H}^{(4)},\\
\label{eq:Bdef}
B &= \tfrac12\Big( \big({\rm tr}\, \mathfrak{H}^{(4)}\big)^2 - {\rm tr}\, \big({\mathfrak{H}^{(4)}}\big)^2\Big),\\
\label{eq:Cdef}
C &= -\tfrac16 \Big( 
\big({\rm tr}\, \mathfrak{H}^{(4)}\big)^3 - 3\, {\rm tr}\, \big({\mathfrak{H}^{(4)}}\big)^2 \big({\rm tr}\, \mathfrak{H}^{(4)}\big)
+ 2\,{\rm tr}\, \big({\mathfrak{H}^{(4)}}\big)^3\Big),\\
\label{eq:Ddef}
D &= \det {\mathfrak{H}^{(4)}},
\end{align}
\textsl{and where} 
\begin{align}\label{eq:Hfour}
& 
\mathfrak{H}^{(4)}= \\ \notag
&\hspace{-2.5truecm}  
{\begin{pmatrix}
{[\![}1{]\!]} 
&  \frac{1}{\sqrt{3}}\bigl({[\![}2{]\!]}+{[\![}1{]\!]} \bigr)
&  \frac{1}{\sqrt{5}}\bigl({[\![}3{]\!]} + {[\![}2{]\!]} \bigr)  
&  \frac{1}{\sqrt{7}}\bigl({[\![}4{]\!]}+{[\![}3{]\!]} \bigr)   \\
 \frac{1}{\sqrt{3}}\bigl({[\![}2{]\!]}+{[\![}1{]\!]} \bigr)
&  \frac{1}{3}\bigl({[\![}3{]\!]}-2{[\![}1{]\!]} \bigr)
&  \frac{1}{\sqrt{15}}\bigl({[\![}4{]\!]}+{[\![}1{]\!]} \bigr) 
&  \frac{1}{\sqrt{21}}\bigl({[\![}5{]\!]}+{[\![}2{]\!]} \bigr)  \\ 
 \frac{1}{\sqrt{5}}\bigl({[\![}3{]\!]}+{[\![}2{]\!]} \bigr) 
&  \frac{1}{\sqrt{15}}\bigl({[\![}4{]\!]}+{[\![}1{]\!]} \bigr) 
&  \frac{1}{5}\bigl({[\![}5{]\!]}-2({[\![}2{]\!]}+{[\![}1{]\!]}) \bigr) 
&  \frac{1}{\sqrt{35}}\bigl({[\![}6{]\!]}+{[\![}1{]\!]} \bigr)  \\
 \frac{1}{\sqrt{7}}\bigl({[\![}4{]\!]}+{[\![}3{]\!]} \bigr)  & 
 \frac{1}{\sqrt{21}}\bigl({[\![}5{]\!]}+{[\![}2{]\!]} \bigr) & 
  \frac{1}{\sqrt{35}}\bigl({[\![}6{]\!]}+{[\![}1{]\!]} \bigr)  & 
 \frac{1}{7}\bigl({[\![}7{]\!]}-2({[\![}3{]\!]}+{[\![}2{]\!]}+{[\![}1{]\!]}) \bigr)   \\
\end{pmatrix}}\!,
\end{align}
\textsl{with (restoring the dependence on $\varpi$)} ${[\![}n{]\!]}(\varpi):= \frac{\varpi^2}{\varpi^2+n^2}$ \textsl{for} $n\in\NN$.
\smallskip

 Also the explicit rigorous upper bound on $\mathfrak{k}(P,T)$ obtained in \cite{KAYdispersive} can now be evaluated
in elementary closed form as rigorous upper bound on $\mathfrak{h}(\varpi)$, which translates into a rigorous lower bound on
on $L_{\mbox{\tiny{E}}}(\varpi)$.
 Explicitly, we have
\smallskip

\noindent
{\bf Theorem~4}: \textsl{Let $\varpi> 0$ be given. Then} $L_{\mbox{\tiny{E}}}(\varpi)\geq 1/\mathfrak{h}^*(\varpi)$, 
\textsl{where
\begin{equation}\label{eq:hSTAR}
\mathfrak{h}^*(\varpi)
=
  \frac{\varpi^2}{1+\varpi^2} + 2\Big(\big(2^{1+\eps}-1\big)\zeta(1+\eps)\zeta(5-\eps)\Big)^\frac12 \varpi^2,
\end{equation}
with $\eps=0.65$.} 
\smallskip

 Our Theorems~1 -- 4 do not rule out that some lines 
$\mathscr{L}(\lambda,\Omega):=\{(\lambda,\Omega,T)\in\RR_+^3: \lambda= c_1\ \&\ \Omega= c_2\}$ could pierce $\mathscr{S}_{\!c}$ more 
than once, in which case the critical surface would not be a graph over the positive $(\lambda,\Omega)$-quadrant
of the {electron-phonon} model parameters. 
 This would be at odds with the narrative that is expected to hold for the Eliashberg model with Einstein phonons, 
for a multiple piercing would mean 
that there is no unique critical temperature $T_c(\lambda,\Omega)$ for certain $(\lambda,\Omega)$.
 To rigorously confirm the empirical thermodynamic narrative for the Eliashberg model, still \textsl{assuming} the existence
of a continuous transition between normal and superconducting phases, one needs to show that $L_{\mbox{\tiny{E}}}(\varpi)$ depends 
strictly monotonically on $\varpi$.
 Monotonicity for a bounded interval of $\varpi$ values follows from the pertinent result in \cite{KAYdispersive}.
\smallskip

\noindent
{\bf Theorem~5}: 
\textsl{For all $N\in\NN$ the eigenvalues $\mathfrak{h}^{(N)}(\varpi)$ increase strictly monotonically with $\varpi\in[0,\varpi_*)$.
 Moreover, $\varpi_*\geq \sqrt{2}$.
 As a consequence, the map} $\varpi\mapsto L_{\mbox{\tiny{E}}}(\varpi)$ 
\textsl{is strictly monotonic decreasing on $[0,\varpi_*]$, with $\varpi_* \geq \sqrt{2}$.
 Thus the portion of the critical surface $\mathscr{S}_{\!c}$ 
over the region $\{\lambda \geq \lambda_*\}$ in the positive $(\lambda,\Omega)$-quadrant is also a graph, 
yielding the critical temperature $T_c(\lambda,\Omega)$, viz.}

\vspace{-.5truecm}
\begin{equation}
 \mathscr{S}_{\!c}\big|_{\lambda\geq\lambda_*}^{} = 
\big\{(\lambda,\Omega,T)\in\RR_+^3: T = T_c(\lambda,\Omega), \lambda\geq\lambda_* \big\}.
\end{equation}
\textit{Moreover, $T_c(\lambda,\Omega) = \Omega f(\lambda)$, where $f(\lambda)$ is continuous and strictly monotonically increasing 
for $\lambda\geq\lambda_*$. 
 Furthermore, $\lambda_* <0.7670...$.}
\smallskip

 While we have not succeeded in showing that the map $\varpi\mapsto\mathfrak{h}(\varpi)$ is strictly monotonic increasing
for all $\varpi\in\RR_+$, our lower bounds to $\mathfrak{h}(\varpi)$ stated explicitly in Theorem~3 for $N\in\{1,2,3,4\}$
all are strictly monotonic increasing with $\varpi$. 
 This is manifestly obvious only for $\mathfrak{h}^{(1)}(\varpi)$. 
 For $\mathfrak{h}^{(2)}(\varpi)$ this is a consequence of Proposition~9 in \cite{KAYdispersive}.
 For $\mathfrak{h}^{(3)}(\varpi)$ and $\mathfrak{h}^{(4)}(\varpi)$ the monotonicity for $\lambda < \lambda_*$ 
is vindicated through visual inspection of the plots (see below).

 We note that the explicit upper bound (\ref{eq:hSTAR}) on $\mathfrak{h}(\varpi)$ is manifestly strictly monotone increasing with
$\varpi$ on $\RR_+$.

 Our Theorems~1 and 2 reveal that the critical surface $\mathscr{S}_{\!c}$ in the positive $(\lambda,\Omega,T)$-octant is a \textsl{ruled surface}
that maps into a \textsl{critical curve} $\mathscr{C}_c$ in the positive $(\lambda,\varpi)$-quadrant, and that critical
curve is a graph over the positive $\varpi$ axis. 
 By Theorems~3 and~4 in concert, that graph $L_{\mbox{\tiny{E}}}(\varpi)$ is sandwiched between $1/\mathfrak{h}^*(\varpi)$
(explicit lower bound) and $1/\mathfrak{h}^{(N)}(\varpi)$ for any $N\in\NN$ (a decreasing sequence of upper bounds).
\smallskip

 Furthermore, by Theorem~5 that critical curve defines a unique critical temperature $T_c(\lambda,\Omega)=\Omega f(\lambda)$ at least for
all $\lambda >\lambda_*$, with $\lambda_*<0.7670...$.
 By their strict monotonic dependence on $\varpi$, also 
our explicit upper bound $\mathfrak{h}^*(\varpi)$ on $\mathfrak{h}(\varpi)$ can be inverted to yield an upper 
\textsl{critical-temperature bound} $T_c^*(\lambda,\Omega) = \Omega f^*(\lambda)$,
and our explicit lower bounds $\mathfrak{h}^{(N)}(\varpi)$, $N\in\{1,2,3,4\}$, on $\mathfrak{h}(\varpi)$ can be inverted to yield 
lower \textsl{critical-temperature bounds} $T_c^{(N)}(\lambda,\Omega)=\Omega f^{(N)}(\lambda)$, $N\in\{1,2,3,4\}$.
 Only $f^*(\lambda)$ and $f^{(1)}(\lambda)$ can be expressed in closed form, though.
 Yet we have convenient explicit parameter representations of $f^{(N)}(\lambda)$ for all $N\in\{1,2,3,4\}$.

 We state this as 
\smallskip

\noindent
{\bf Corollary~1}:
 \textsl{For $\lambda >1$ we have
\begin{equation}\label{eq:TcONE}
T_c^{(1)}(\lambda,\Omega) 
=
\tfrac{\Omega}{2\pi}\sqrt{\lambda -1}, 
\end{equation}
while for $\lambda >0$ we have
\begin{equation}\label{eq:TcSTAR}
T_c^{*}(\lambda,\Omega) 
=
\tfrac{\Omega}{2\pi}\sqrt{\tfrac12\Big(\lambda(1+b) -1 + \sqrt{\big(\lambda(1+b)-1\big)^2+4b\lambda}\Big)}
\end{equation}
with $b:=2\Big(\big(2^{1+\eps}-1\big)\zeta(1+\eps)\zeta(5-\eps)\Big)^\frac12$ and $\eps=0.65$.}

\textsl{For $T_c^{(N)}(\lambda,\Omega)$ with $N\in\{2,3,4\}$ we have
\begin{equation}\label{eq:TcLOWERboundsNoneTOfour}
T_c^{(N)}(\lambda,\Omega) = \Omega f^{(N)}(\lambda),
\end{equation}
where $\lambda\mapsto f^{(N)}(\lambda)$ are the special cases $N\in\{2,3,4\}$ of the curves 
$\tilde{\mathscr{C}}_c^{(N)}$ in the positive $(\lambda,T/\Omega)$-quadrant that $\forall\, N\in\NN$ are given by 
\begin{equation}\label{eq:tildeCcritPARAM}
\tilde{\mathscr{C}}_c^{(N)}= 
\left\{\left(\lambda,\tfrac{T}{\Omega}\right)\in \RR_+^2: \lambda = \tfrac{1}{\mathfrak{h}^{(N)}(\varpi)}\ \&\ 
\tfrac{T}{\Omega}= \tfrac{1}{2\pi \varpi} \right\};\quad N\in \NN.
\end{equation}
 If $N\in\{1,2,3,4\}$ then $\tilde{\mathscr{C}}_c^{(N)}$ is a graph over the interval $[\lambda_N,\infty)$
on the $\lambda$ axis, where $\lambda_N$ is the endpoint of 
$\tilde{\mathscr{C}}_c^{(N)}$ on the $\lambda$ axis.}
\smallskip

 We conjecture that all $\tilde{\mathscr{C}}_c^{(N)}$ are such graphs for general $N\in\NN$.
 So far, for general $N\in\NN$, our Theorem~5 guarantees that each $\tilde{\mathscr{C}}_c^{(N)}$ is a graph
over $[\max\{\lambda_*,\lambda_N\},\infty)$, for some $\lambda_*<0.7670...$.
 Moreover, below we show that each $\tilde{\mathscr{C}}_c^{(N)}$ is asymptotic to a graph over a right neightborhood
of $\lambda_N$, with $\lambda_N\searrow 0$ given in (\ref{eq:lambdaN}).

 Besides knowing the best explicitly available ``sandwiching bounds'' on $T_c(\lambda,\Omega)$, 
it is of interest to plot all these bounds in a single diagram to get some visual impression of the speed of convergence;
recall that we only know that the sequence $L_{\mbox{\tiny{E}}}^{(N)}(\varpi)$ converges downward to $L_{\mbox{\tiny{E}}}(\varpi)$ when $N\to\infty$, 
but don't know how fast --- analogously for the lower critical-temperature bounds $T_c^{(N)}(\lambda,\Omega)$ 
when $\lambda>\max\{\lambda_N,0.7671\}$, for which these bounds are well-defined functions of $\Omega$ and $\lambda$
through inversion of $L_{\mbox{\tiny{E}}}^{(N)}(\varpi)$.
 In this spirit, the lower bounds $T_c^{(N)},N\in\{1,2,3,4\}$, the upper bound $T_c^*$ on any $T_c$ 
stated in Corollary~1, and the conjectured upper bound $T_c^{\sim}$ of Conjecture~1 below, 
are displayed in Fig.~1 as functions of $\lambda$. 

\begin{figure}[H]
 \centering
\includegraphics[width=0.7\textwidth,scale=0.7]{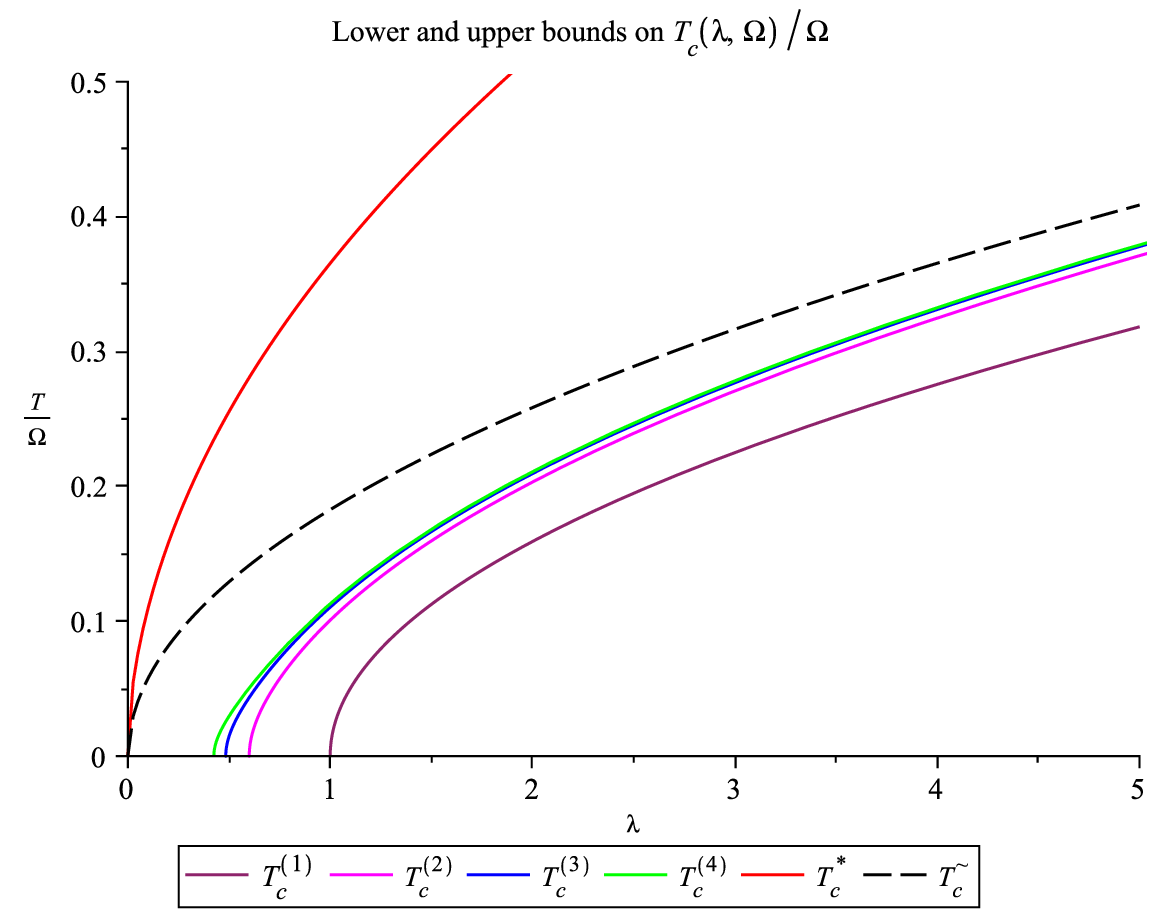} 
\vspace{-10pt}
\caption{\label{fig:TcBOUNDSvsLAMBDA}
\footnotesize
 Shown are the graphs of the maps $\lambda\mapsto T_c^{(N)}(\lambda,\Omega)/\Omega$ for $N\in\{1,2,3,4\}$,
the graph of the map $\lambda\mapsto T_c^*(\lambda,\Omega)/\Omega$,
and the graph of the map $\lambda\mapsto T_c^{\sim}(\lambda,\Omega)/\Omega$ of Conjecture~1.
 All displayed curves are plots of $\lambda$ as given by
some explicitly computed elementary functions of $T/\Omega$.}
\end{figure}
\vspace{-15pt}

 It is obvious from Figure~1 that our upper bound on $T_c(\lambda,\Omega)$ is not optimal; yet it agrees with the asymptotic 
behavior $T_c(\lambda,\Omega) \sim C \Omega\sqrt{\lambda}$ for large enough $\lambda$, given $\Omega$, 
though with $C\approx 2.034 C_\infty$, where $C_\infty = 0.1827262477...$ is the optimal constant. 

 Further visual inspection of Figure~1 reveals that
the sequence of lower bounds $T_c^{(N)}(\lambda,\Omega)$ on $T_c(\lambda,\Omega)$ 
appears to converge upward very rapidly to some limiting curve $T_c(\lambda,\Omega)=\Omega f(\lambda)$ when $\lambda> 0.7$ (say).
 For then the gap between the $T_c^{(3)}$ and $T_c^{(4)}$ curves is so small that the line width of the plotted curves fills it.

 On the other hand, when $\lambda$ is less than $\approx 0.5$, then the gap between these two curves becomes clearly visible.
 In fact, near the $\lambda$ axis convergence is slow.
 The sequence of upper bounds $L_{\mbox{\tiny{E}}}^{(N)}(\varpi)$ to $L_{\mbox{\tiny{E}}}(\varpi)$ meets the $\lambda$ axis at explicitly 
computable locations $\lambda_N$ that converge slowly to $0$ like $1/\ln N$ as $N\to\infty$.

 More precisely, we have:
\smallskip

\noindent
{\bf Theorem~6}:\hspace{-3pt}
\textsl{The eigenvalues $\mathfrak{h}^{(N)}(\varpi)$ are analytic about $\varpi=\infty$, with}
\vspace{-0.8truecm}

\begin{equation}\label{eq:CcurveNasymp}
\mathfrak{h}^{(N)}(\varpi)
= \mathfrak{h}^{(N)}(\infty) - B_N \tfrac{1}{\varpi^2} + \mathcal{O}\big(\tfrac{1}{\varpi^{4}}\big)
\end{equation}
\vspace{-0.5truecm}

\textsl{where}
\vspace{-0.5truecm}

\begin{equation}\label{eq:lambdaN}
\mathfrak{h}^{(N)}(\infty) 
= -1 + 2{\textstyle\sum\limits_{n=0}^{N-1} \frac{1}{2n+1}}
\end{equation}
\vspace{-0.5truecm}

\textsl{and}
\vspace{-0.5truecm}

\begin{equation}\label{eq:BN}
B_N = 
\frac{\sum\limits_{n=0}^{N-1}\sum\limits_{m=0}^{N-1}\frac{(n-m)^2 + (n+m+1)^2}{(2n+1)(2m+1)}
-\sum\limits_{\ell=0}^{N-1}\frac{2}{(2\ell+1)^2}\sum\limits_{k=1}^{\ell}k^2}{\sum\limits_{j=0}^{N-1}\frac{1}{2j+1}}
\geq 1.
\end{equation}
\vspace{-0.6truecm}

\textsl{Thus, as $\varpi\to\infty$ the $N$-th upper approximation}
 $L_{\mbox{\tiny{E}}}^{(N)}(\varpi)$ \textsl{to the critical curve given by} $\varpi\mapsto L_{\mbox{\tiny{E}}}(\varpi)$ 
\textsl{converges downward to $\lambda_N = 1/\mathfrak{h}^{(N)}(\infty)$.
 Moreover, both $\mathfrak{h}^{(N)}(\infty)$ and $B_N$ are strictly monotonically increasing with $N$, diverging to $\infty$
as $N\to\infty$.}
\smallskip

 Only the existence of the $\mathfrak{h}^{(N)}(\infty)$ follows as special case of the analogous result about the $T\to 0$ limit
of the eigenvalues $\mathfrak{k}^{(N)}(P,T)$ of the Eliashberg model with dispersive phonons that we proved in \cite{KAYdispersive}.
 Theorem~6 in full therefore will be proved in this paper. 
 It establishes that the $L_{\mbox{\tiny{E}}}^{(N)}(\varpi)$ are asymptotic to strictly
monotonically decreasing functions when $\varpi\sim\infty$.
 Thus, in the vicinity of the point $\big(\lambda_N,0\big)$, 
the critical curve $\tilde{\mathscr{C}}_c^{(N)}$ in the positive $(\lambda,T/\Omega)$-quadrant is
also asymptotic to a graph over some small interval on the $\lambda$ axis to the right of $\lambda_N$,
there defining $T_c^{(N)}(\lambda,\Omega)$.
 More precisely, we have
\smallskip

\noindent
{\bf Corollary~2}:
\textsl{As $\lambda$ approaches $\lambda_N$ from the right, we have}
\begin{equation}\label{eq:TcNlambdaN}
T_c^{(N)}(\lambda,\Omega) 
\sim
\tfrac{\Omega}{2\pi}\sqrt{\tfrac{1}{B_N}\left(\tfrac{1}{\lambda_N} - \tfrac1\lambda \right)} , 
\quad\mbox{for}\quad N\in\NN.
\end{equation}

 Together with our upper bound $T_c^*(\lambda,\Omega)$ this proves that the continuous critical curve $\tilde{\mathscr{C}}_c$ that
divides the positive $(\lambda,T/\Omega)$-quadrant into simply connected normal and superconducting
regions, and which is a graph over the $T/\Omega$ axis, starts at $(0,0)$.
 The upper bound on $T_c$ guarantees that $(0,0)$ is the only point of the critical curve on the $T/\Omega$ axis.
 The upper bound on $T_c$ in concert with any of the lower bounds $T_c^{(N)}$, $N\in\{1,2,3,4\}$, in turn proves that 
$\tilde{\mathscr{C}}_c$ goes to $(\infty,\infty)$, asymptotically for $\lambda\sim\infty$ bounded above and below 
$\propto\Omega\sqrt{\lambda}$.

 The lower bound on $L_{\mbox{\tiny{E}}}(\varpi)$, and thus the upper bound on $T_c(\lambda,\Omega)$, can certainly be improved,
yet it is a challenging task to improve it to the precision that is suggested by a small $\varpi$ analysis of the operators
$\mathfrak{H}^{(N)}(\varpi)$, and the apparent rapid convergence of the sequence of eigenvalues $\mathfrak{h}^{(N)}(\varpi)$
for small $\varpi$.
 Small $\varpi$ analysis yields the same result as the large-$T$ analysis of the dispersive phonons paper \cite{KAYdispersive}
in the special case $P(d\omega) = \delta(\omega-\Omega)d\omega$, viz.
\smallskip

\noindent
{\bf Theorem~7}: \textsl{The eigenvalues $\mathfrak{h}^{(N)}(\varpi)$ are analytic about $\varpi=0$, with}

\vspace{-0.5truecm}
\begin{equation}\label{eq:CcurveNasympSMALL}
\mathfrak{h}^{(N)}(\varpi)
= \mathfrak{g}^{(N)}(2) \varpi^2  - \bigl\langle \mathfrak{G}^{(N)}(4)\bigr\rangle^{}_{\!2}\varpi^4 + \mathcal{O}\big(\varpi^{6}\big)
\end{equation}
\textsl{where $\mathfrak{g}^{(N)}(2)$ is the largest eigenvalue for the $N$-Matsubara frequency approximation 
to the operator $\mathfrak{G}(\gamma)$ of the $\gamma$ model at $\gamma=2$, and where
{$\langle \mathfrak{G}^{(N)}(4)\rangle^{}_{\!2}>0$} denotes the quantum-mechanical expected value of 
the $N$-Matsubara frequency approximation to the operator $\mathfrak{G}(\gamma)$ at $\gamma=4$,
taken with the $N$-frequency optimizer of the $\gamma$  model at $\gamma=2$.}

\noindent
{\bf Corollary~3}:
\textsl{The $N$-Matsubara frequency approximation $\tilde{\mathscr{C}}_c^{(N)}$ to the  critical curve
$\tilde{\mathscr{C}}_c$ in the positive $(\lambda,\frac{T}{\Omega})$-quadrant is asymptotic to a graph over the asymptotic
region $\lambda\sim\infty$ of the $\lambda$ axis, given by}
\begin{equation}\label{eq:TcNlambda}
\hspace{-0.5truecm}
T_c^{(N)}(\lambda,\Omega) 
\sim
\frac{\Omega}{2\pi}\frac{1}{\sqrt{\frac12\frac{\mathfrak{g}^{(N)}(2)}{\left\langle \mathfrak{G}^{(N)}(4)\right\rangle^{}_{\!2}}
\biggl(1-\sqrt{1-4\frac{\left\langle \mathfrak{G}^{(N)}(4)\right\rangle^{}_{\!2}}{\mathfrak{g}^{(N)}(2)^2}\frac1\lambda }\,\biggr)}}.
\end{equation}
\textsl{This result also holds when $N\to\infty$ (with the superscripts ${}^{(N)}$ purged).}
\vspace{-0.5truecm}

 For large $\lambda$ the r.h.s.(\ref{eq:TcNlambda}) can be expanded to yield 
$T_c^{(N)}(\lambda,\Omega) \sim \frac{\Omega}{2\pi}\sqrt{\mathfrak{g}^{(N)}(2)\lambda}$, with $N\in\NN$.
 By a simple convexity estimate, r.h.s.(\ref{eq:TcNlambda})$\leq \frac{\Omega}{2\pi}\sqrt{\mathfrak{g}^{(N)}(2)\lambda}$,
so the asymptotic expression $\frac{\Omega}{2\pi}\sqrt{\mathfrak{g}^{(N)}(2)\lambda}$ is an upper bound on 
$T_c^{(N)}(\lambda,\Omega)$ for large enough $\lambda$ that is asymptotically sharp as $\lambda\sim\infty$.
 Moreover, in \cite{KAYgamma} we showed that $\mathfrak{g}^{(N)}(2)$ converges upward to $\mathfrak{g}(2)$.
 Furthermore, as noted earlier, each $T_c^{(N)}(\lambda,\Omega)$ vanishes for $\lambda\leq \lambda_N$, while $\sqrt{\lambda}>0$ 
for all $\lambda$. 
 All the above, plus the rapid convergence for $\lambda>0.77$ of the $T_c^{(N)}(\lambda,\Omega)$ with $N$ discernible in Fig.~1
now suggests

\noindent
{\bf Conjecture~1}: \textsl{There is a critical temperature $T_c(\lambda,\Omega) >0$ which 
for all $\Omega> 0$ and $\lambda >0$ is bounded above by $T_c(\lambda,\Omega)< T_c^{\sim}(\lambda,\Omega)$, with
\begin{equation}\label{eq:CONJbound}
T_c^{\sim}(\lambda,\Omega) := \tfrac{\Omega}{2\pi}\sqrt{\mathfrak{g}(2)\lambda};
\end{equation}
here, $\mathfrak{g}(2)$ is the spectral radius of the operator $\mathfrak{G}(\gamma)$ at $\gamma=2$.}

 For the numerical approximation of $\frac{1}{2\pi}\sqrt{\mathfrak{g}(2)}$ to 10 significant decimal places,
see the comments in the introduction of \cite{KAYgamma}.

\noindent
{\bf Remark~1}: \textsl{{We remark that the asymptotic behavior $T_c(\lambda,\Omega) \sim C \Omega\sqrt{\lambda}$ for 
large enough $\lambda$, given $\Omega$, 
 was anticipated in \cite{AllenDynes}, though based on nonrigorous arguments; see their eq.(22).
  Incidentally, in \cite{AllenDynes} the value $C=0.182$ was stated to be
computed with a 64 Matsubara mode approximation to the linearized Eliashberg gap equation.
 Apparently their computation was not very accurate, for the rounded approximation to three significant digits 
reads $C=0.183$, and 0.183 is already the rounded value of the closed form approximation with merely four
Matsubara modes obtained in our paper.}}

 The next figure shows the large-$\lambda$ behavior of our four lower bounds,
and of the conjectured global upper bound, on  $T_c(\lambda,\Omega)$. 
 \vspace{-12pt}

\begin{figure}[H]
 \centering
\includegraphics[width=0.7\textwidth,scale=0.7]{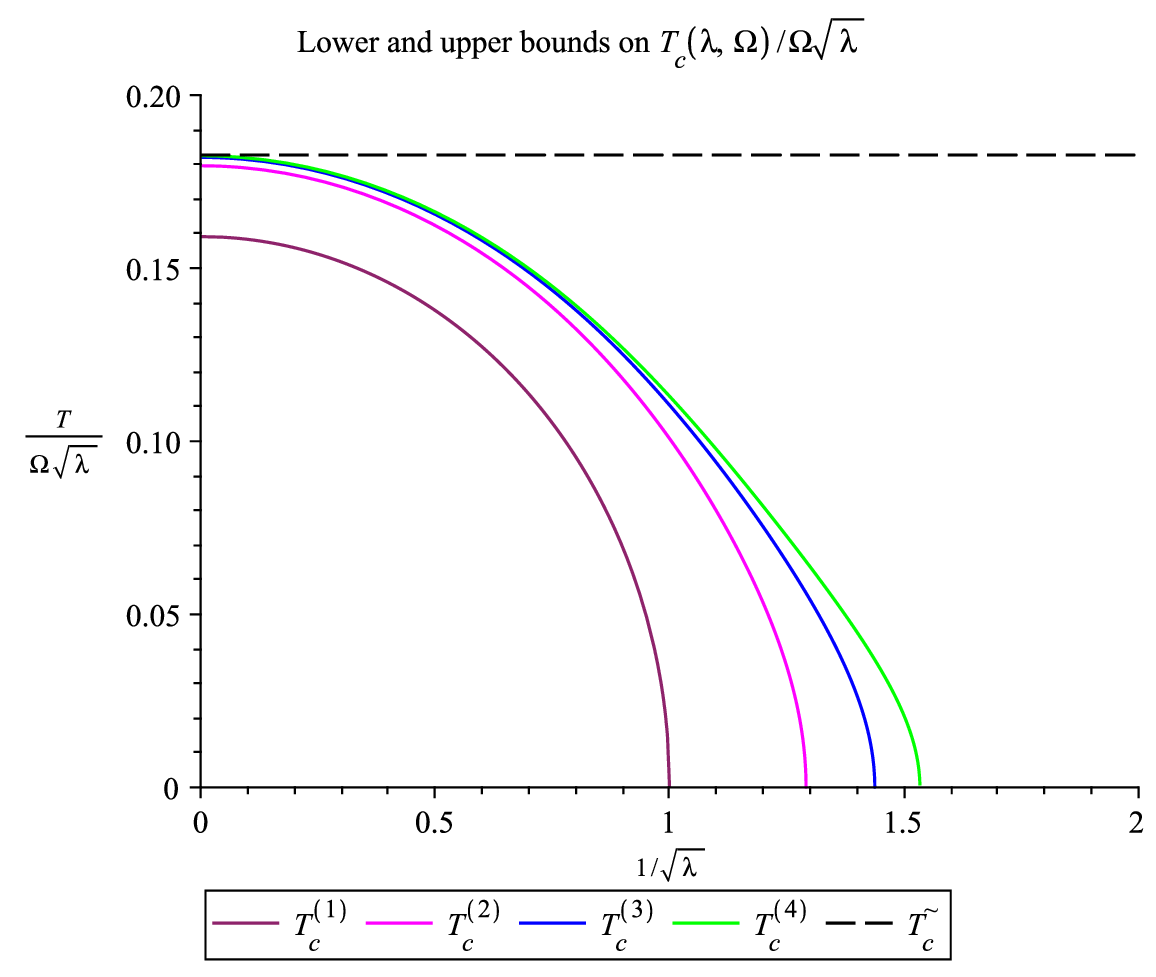} 
\caption{\label{fig:TcLOWERbounds123vsOMEGA}
\footnotesize
Shown are the graphs of the maps ${1}/{\sqrt{\lambda}}
\mapsto T_c^{(N)}(\lambda,\Omega)\big/{\Omega\sqrt{\lambda}}$ for $N\in\{1,2,3,4\}$, with the 
$T_c^{(N)}(\lambda,\Omega)$ 
our explicitly computed members of the strictly monotonically increasing sequence of lower bounds to $T_c(\lambda,\Omega)$.
 The third and fourth approximates virtually agree with each other if $\frac{1}{\sqrt{\lambda}} <0.5$,
and only minute discrepancies are visible when $0.5<\frac{1}{\sqrt{\lambda}} < 1$, indicating the rapid 
convergence of our lower bounds to the exact result when $\lambda >1$. 
 All four lower bounds hit zero, the $N$-th one at $\frac{1}{\sqrt{\lambda_N}} \geq 1$.
 The conjectured global upper bound $T_c^\sim(\lambda,\Omega)$ (dashed line) is visibly seen to be asymptotically exact
(as proved), yet not very accurate away from the asymptotic regime $\lambda\sim\infty$.
 Our rigorous bound (not shown) would be an almost horizontal line twice as high as the dashed one.
}
\end{figure}
\vspace{-12pt}

 The dashed horizontal line in Fig.~2 highlights the
asymptotic connection with the $\gamma$ model, made precise in the following proposition. 

\noindent
{\bf Proposition~1}: 
\textit{We have
\begin{equation}
\lim_{\lambda\to\infty} \tfrac{1}{\sqrt{\lambda}} f(\lambda) = \tfrac1g T_c(g,2),
\end{equation}
where $T_c(g,\gamma)$ is the critical temperature of the $\gamma$ model.
{Numerically, a 200 mode approximation yields $\tfrac1g T_c(g,2) = 0.1827262477...$.}}

 We now turn to the verification of our results. 

\section{Verification of the main results}\label{sec:VERIFY}

  We will present only the proofs of those results that are not special cases of the results that we proved in \cite{KAYdispersive}.
 For all other results we confine ourselves to stating the simplifications in the proofs of our more general results of \cite{KAYdispersive}
for dispersive phonons.

 The results stated in the previous section are based on the linear stability analysis of the normal state in Eliashberg theory,
carried out in \cite{KAYdispersive}, specialized to the non-dispersive limit $P(d\Omega) = \delta(\omega-\Omega)d\omega$.
 Thus, we again work with a normalized version of a functional given in \cite{YuzAltPRB}, 
known as the \textsl{condensation energy} of Eliashberg theory [the difference between the grand (Landau) potentials of 
the superconducting and normal states], using units where Boltzmann's constant $k_{\mbox{\tiny{B}}}=1$ and the reduced 
Planck constant $\hbar =1$.
 Its Bloch spin chain representation reads (cf. \cite{KAYgamma}, eq.(32))
\begin{align}\label{eq:H}
 {H}({\bf S}|{\bf N}) : = &2\pi \sum\limits_{n} \omega_n {\bf N}_0\cdot\big({\bf N}_n -{\bf S}_n \big)
  \\
\notag & + \pi^2 {T}
\sum\!\sum\limits_{\hskip-0.4truecm n\neq m}
\lambda_{n,m}^{} \left({\bf N}_n\cdot {\bf N}_m -{\bf S}_n\cdot{\bf S}_m \right),
\end{align}
where $\lambda_{n,m}^{}$ is a (dimensionless) positive spin-pair interaction kernel, chosen below, and
where the summations here run over ${\ZZ}$.
 In (\ref{eq:H}), ${\bf N}\in ({\SP}^1)^{\ZZ}$ is the Bloch spin-chain associated with the {\it{normal state}} 
of the Migdal--Eliashberg theory, having $n$-th spin given by ${\bf N}_n :=-{\bf N}_0\in {\SP}^1\subset{\RR}^2$ for $n < 0$
and ${\bf N}_n:={\bf N}_0$ for $n\geq 0$.
 Any other Bloch spin chain ${\bf S}\in ({\SP}^1)^{\ZZ}$ \textsl{admissible} in (\ref{eq:H}) has to 
satisfy the asymptotic conditions that, sufficiently fast,
${\bf S}_n \to {\bf N}_n$ when $n\to \infty$ and when $n\to - \infty$, where ${\bf S}_n \in {\SP}^1\subset {\RR}^2$ with $n\in{\ZZ}$ 
denotes the $n$-th spin in the spin chain ${\bf S}$, and where ``sufficiently fast'' is explained below.
{In addition, \textsl{admissible} spin chains must satisfy the symmetry relationship that for all $n\in{\ZZ}$, 
 ${\bf N}_0\cdot {\bf S}_{-n} = - {\bf N}_0\cdot{\bf S}_{n-1}$ and 
 ${\bf K}_0\cdot {\bf S}_{-n} =  {\bf K}_0\cdot{\bf S}_{n-1}$, 
where ${\bf K}_0\in {\SP}^1\subset {\RR}^2$ is an arbitrary vector perpendicular to ${\bf N}_0$.}

 Since in this paper we study the Eliashberg model with Einstein phonons of frequency $\Omega$,
the spin-pair interaction kernel $\lambda_{n,m}^{}$ reads (cf. \cite{KAYgamma}, eq.(11))
\begin{equation}\label{eq:lambdaHOLSTEIN}
\lambda_{n,m}^{} := \lambda \frac{\Omega^2}{\Omega^2+(\omega_n-\omega_m)^2};
\end{equation}
here, $\lambda$ is the dimensionless electron-phonon coupling constant of the theory.
 Note that our $\lambda$ is the \textit{standard} (renormalized) dimensionless electron-phonon coupling constant 
of the Eliashberg theory; cf. \cite{AllenDynes}.
 Note also that r.h.s.(\ref{eq:lambdaHOLSTEIN}) is the special case $P(d\omega)=\delta(\omega-\Omega)d\omega$ 
of the r.h.s.(7) in \cite{KAYdispersive}.

 Since $\omega_n-\omega_m = (n-m)2\pi T $, it has also become customary to use the notation 
$\lambda(n-m)$ instead of $\lambda_{n,m}^{}$, and to write $\lambda=\lambda(0)$. 
 In order to avoid any ambiguous statements, 
we will use $\lambda$  exclusively to mean the coupling constant (\ref{eq:lambdaISgSQoverOMsq}), 
and not (as sometimes done in the superconductivity literature) as abbreviation for the map 
$j\mapsto \lambda(j)$, with $j\in\ZZ$.

 Incidentally, (\ref{eq:lambdaHOLSTEIN}) can also be rewritten as
\begin{equation}\label{eq:Vphonon}
\lambda_{n,m}^{}  =\frac{g^2}{\Omega^2+(\omega_n-\omega_m)^2}
\end{equation}
(cf. eq.(9) in \cite{KAYgamma}), 
and then $\lambda$ is given in terms of $g$ and $\Omega$ as
\begin{equation}\label{eq:lambdaISgSQoverOMsq}
\lambda=\frac{g^2}{\Omega^2}
\end{equation}
(cf. eq.(10) in \cite{KAYgamma}).
 Using this representation (\ref{eq:Vphonon}) of $\lambda_{n,m}$, and taking the limit $\Omega\searrow 0$ while keeping $g$ fixed,
one obtains the condensation energy functional for the $\gamma$ model at $\gamma=2$.
 Subsequently replacing $g^2\to g^\gamma$ and $(\omega_n-\omega_m)^2\to |\omega_n-\omega_m|^\gamma$ with $\gamma>0$ one obtains
the condensation energy functional for the $\gamma$ model discussed in \cite{KAYgamma}.

 We now complete our definition of \textsl{admissibility} of a spin chain ${\bf S}$ to mean that after expressing the summations over 
negative Matsubara frequencies by summations over positive ones, as per the symmetry relationship 
 ${\bf N}_0\cdot {\bf S}_{-n} = - {\bf N}_0\cdot{\bf S}_{n-1}$ and 
${\bf K}_0\cdot {\bf S}_{-n} =  {\bf K}_0\cdot{\bf S}_{n-1}$ for all $n\in{\ZZ}$,
the resulting series resulting from the sum and double sum in (\ref{eq:H}), viz. (\ref{eq:K}), converge absolutely.

 Having introduced the condensation energy functional for the Eliashberg model with Einstein phonons, we can now rephrase
the ``thermodynamic narrative'' of the introduction in a precise manner.
\smallskip

\noindent
{\bf Conjecture~2}: \textsl{
There is a critical temperature $T_c >0$, depending on $\lambda> 0$ and $\Omega >0$, 
such that for temperatures  $T\geq T_c(\lambda,\Omega)$, the spin chain of the normal state $\mathbf{N}$ is the unique
minimizer of $H({\mathbf S}|{\mathbf N})$, whereas at temperatures $T<T_c(\lambda,\Omega)$ a spin chain ${\bf S}\neq {\bf N}$ 
for a {\it{superconducting phase}} minimizes $H({\mathbf S}|{\mathbf N})$ uniquely 
up to an irrelevant gauge transformation (fixing of an overall phase).
 Moreover, the phase transition at $T_c$ from normal to superconductivity is continuous.} 
\smallskip

 Conjecture~2, if confirmed, implies that the normal (metallic) state is linearly stable against small
perturbations toward the superconducting region when $T>T_c(\lambda,\Omega)$, and unstable when $T<T_c(\lambda,\Omega)$.
 This is the stability criterion we will study in the following, by
expanding ${H}({\bf S}|{\bf N})$ about ${\bf N}$ to second order in the perturbations and study its 
minimization over the set of normalized perturbations.

 For this investigation it is prudent to first rewrite (\ref{eq:H}) into a more convenient format, following
\cite{KAYgamma} and \cite{KAYdispersive}.
 First of all, the symmetry relationship
 ${\bf N}_0\cdot {\bf S}_{-n} = - {\bf N}_0\cdot{\bf S}_{n-1}$
and 
{${\bf K}_0\cdot {\bf S}_{-n} =  {\bf K}_0\cdot{\bf S}_{n-1}$}
 for all $n\in{\ZZ}$ allows us
to work with effective spin chains ${\bf S}\in  ({\SP}^1)^{{\NN}_0}$, with ${\NN}_0:={\NN}\cup\{0\}$.
 The summations over ${\ZZ}$ can therefore be rewritten in terms of summations over ${\NN}_0:={\NN}\cup\{0\}$.
 Second, the restriction that the vectors ${\bf S}_n$ are in $\SP^1$ is implemented
by introducing an angle $\theta_n\in {\RR}/(2\pi{\ZZ})$ ($= [0,2\pi]$ with $2\pi$ and $0$ identified)
defined through ${\bf N}_0\cdot{\bf S}_n =: \cos \theta_n$ for all\footnote{If one also
introduces angles for spins with negative suffix by defining ${\bf N}_0\cdot{\bf S}_n =: \cos \theta_n$ for 
all $n\in -{\NN}$, a sequence of angles with non-negative 
suffix yields the angles with negative suffix as $\theta_{-1}=\pi-\theta_0$, $\theta_{-2}=\pi-\theta_1$, etc.,
thanks to the symmetry of ${\bf S}\in ({\SP}^1)^{{\ZZ}}$ with respect to the sign switch of the Matsubara frequencies.}
 $n\in {\NN}_0$.
 Setting $ {H}({\bf S}|{\bf N}) =: 4\pi^2 T K(\Theta)$ with $\Theta:=(\theta_n)^{}_{n\in{\NN}_0}$ yields
\begin{align}\label{eq:K}
\hspace{-1truecm}
K_{\mbox{\rm{\tiny{E}}}}(\Theta) = & {\textstyle\sum\limits_n} \biggl[
\big(2 n + 1\big) \big(1-\cos\theta_n \big)
- \lambda\frac{\varpi^2}{2}
\frac{1-\cos\big(2\theta_n\big) }{ \varpi^2 +(2n+1)^2}\biggr]
  \\
\notag 
\hspace{-1truecm} &
+ \lambda\frac{\varpi^2}{2}
{\textstyle\sum\!\sum\limits_{\hskip-0.4truecm n\neq m}}\biggl[\frac{1-\cos\big(\theta_n-\theta_m\big) }{ \varpi^2 + (n-m)^2}
- \frac{1-\cos\big(\theta_n+\theta_m\big) }{ \varpi^2 +(n+m+1)^2}\biggr];
\end{align}
here, the summations run over ${\NN}_0$, and $\varpi: = {\Omega}/2\pi{T} > 0$.
 The functional $K_{\mbox{\rm{\tiny{E}}}}(\Theta)$ stated in (\ref{eq:K}) is the special case $P(d\omega)=\delta(\omega-\Omega)d\omega$ 
of the functional $K(\Theta)$ given in  eq.(34) of \cite{KAYdispersive}.
 The variations of $K_{\mbox{\rm{\tiny{E}}}}(\Theta)$ w.r.t. $\Theta$ yield a {non-linear} Euler--Lagrange equation
for any stationary point $\Theta^s$ of $K_{\mbox{\rm{\tiny{E}}}}(\Theta)$; viz., $\forall n\in{\NN}_0$:
\begin{equation}\label{eq:EL}
\hspace{-0.8truecm}
\big( 2n + 1 \big) \sin\theta_n^s = \lambda{\textstyle\sum\limits_{m \geq 0}} 
\biggl[\frac{\varpi^2\sin\big(\theta_n^s+\theta_m^s\big)}{\varpi^2 +(n+m+1)^2 }- 
\frac{\varpi^2\sin\big(\theta_n^s-\theta_m^s\big) }{\varpi^2 + (n-m)^2}\biggr].
\end{equation}

 In the following we shall omit the superscript ${}^s$ from $\Theta^s$.

 The system of equations (\ref{eq:EL}) has infinitely many solutions when the $\theta_n$ are allowed to take
values in $[0,2\pi]$, restricted only by the asymptotic condition that $\theta_n\to 0$ rapidly enough when $n\to\infty$; 
see \cite{YuzAltPRB}. 
 However, we here are only interested in solutions that are putative minimizers of $H({\mathbf S}|{\mathbf N})$,
i.e. of $K_{\mbox{\rm{\tiny{E}}}}(\Theta)$.
 In \cite{YuzAltPRB} it was shown that a sequence $\Theta=(\theta_n)^{}_{n\in{\NN}_0}$ that minimizes 
$K_{\mbox{\rm{\tiny{E}}}}(\Theta)$, must have $\Theta\in [0,\frac\pi2]^{{\NN}_0} =:S$; i.e.,
all\footnote{Alternatively, all $\theta_n\in[-\frac\pi2,0]$; these choices are gauge equivalent.}
 $\theta_n\in[0,\frac\pi2]$.
 The normal state corresponds to the sequence of angles $\underline\Theta:=(\theta_n =0)^{}_{n\in{\NN}_0}$.
 This trivial solution of (\ref{eq:EL}) manifestly exists for all $\lambda>0$ and $\varpi> 0$.
 We note that $K_{\mbox{\rm{\tiny{E}}}}(\underline\Theta)= 0 = H(\mathbf{N}|\mathbf{N})$. 

\subsection{\hspace{-10pt}Linear stability analysis of the normal state}\label{sec:STAB}

 At last we are ready to inquire into the question of its linear stability versus its instability against modes $\Theta\in S$ for 
which $K_{\mbox{\rm{\tiny{E}}}}(\Theta)$ is well-defined.
 We will show that for all $\varpi> 0$ there is a unique $\lambda = L_{\mbox{\tiny{E}}}(\varpi)>0$
such that the trivial solution $\underline\Theta$ is linearly stable for $\lambda<L_{\mbox{\tiny{E}}}(\varpi)$, 
but unstable against perturbations toward the superconducting region for $\lambda>L_{\mbox{\tiny{E}}}(\varpi)$.
 Moreover, we formulate in detail the variational principle that directly characterizes $L_{\mbox{\tiny{E}}}(\varpi)$.
 This will establish Theorems~1 and~2.

 For the linear stability analysis one needs 
$K_{\mbox{\rm{\tiny{E}}}}(\Theta)$ expanded about $\Theta=\underline\Theta$ to second order in $\Theta$.
 This yields a quadratic form that is the special case  $P(d\omega)=\delta(\omega-\Omega)d\omega$ of eq.(36) of \cite{KAYdispersive}, viz.
\begin{align}\label{eq:Qsimplified}
\hspace{-1truecm}
K_{\mbox{\rm{\tiny{E}}}}^{(2)}(\Theta) = &\, 
{\sum\limits_n} \biggl[
\frac{2 n + 1}{2} 
             - \lambda\biggl(\frac12 \frac{\varpi^2}{ \varpi^2 +(2n+1)^2}
 - {\sum\limits_{k=1}^{n}} \frac{\varpi^2}{\varpi^2 + k^2} \biggr)\biggr] \theta_n^2  \\
\notag &
\!\!- \lambda\frac{1}{2}
\sum 
\!\sum\limits_{\hspace{-0.5truecm}n\neq m} 
\theta_n\biggl[\frac{\varpi^2} 
{ \varpi^2 + (n-m)^2} + \frac{\varpi^2}{ \varpi^2 +(n+m+1)^2}\biggr]\theta_m,
\end{align}
which for all $\lambda>0$ and $\varpi >0$ is well-defined on the Hilbert space $\mathcal{H}$ of
sequences that satisfy $\|\Theta\|_{\mathcal{H}}^2:=\sum_{n\geq 0} (2n+1)\theta_n^2 < \infty$.
 If $K_{\mbox{\rm{\tiny{E}}}}^{(2)}(\Theta) \geq 0$ for all $\Theta\in\mathcal{H}$, with ``$=0$'' iff $\Theta = \underline\Theta$, 
then $K_{\mbox{\rm{\tiny{E}}}}(\Theta)>0$ for all $\Theta\neq\underline\Theta$ in a sufficiently small neighborhood of
$\underline\Theta$, which means that the trivial sequence $\underline\Theta$ is a local minimizer of $K_{\mbox{\rm{\tiny{E}}}}(\Theta)$
and thus linearly stable, then.
 If on the other hand there is at least one $\Theta\neq \underline\Theta$ in $\mathcal{H}\cap S$
for which $K_{\mbox{\rm{\tiny{E}}}}^{(2)}(\Theta)<0$, then the trivial sequence $\underline\Theta$ is 
not a local minimizer of $K_{\mbox{\rm{\tiny{E}}}}(\Theta)$ in $\mathcal{H}\cap S$, and therefore unstable 
against perturbations toward the superconducting region.
 The verdict as to linear stability versus instability depends on $\lambda$ and $\varpi$.

 As in \cite{KAYgamma} and \cite{KAYdispersive}, we recast the functional $K_{\mbox{\rm{\tiny{E}}}}^{(2)}(\Theta)$ 
defined on $\mathcal{H}$ as a functional $Q_{\mbox{\rm{\tiny{E}}}}(\Xi)$ defined on $\ell^2({\NN}_0)$.
 For this we note that we can take the square root of the diagonal matrix $\mathfrak{O}$ whose diagonal elements are 
the odd natural numbers.
 Its square root is also a diagonal matrix, and its action on $\Theta$ componentwise is given as
\begin{equation}\label{eq:Dop}
(\mathfrak{O}^{\frac12}\Theta)_n =  \sqrt{2n + 1}\; \theta_n =: \xi_n.
\end{equation}
 Since $\Theta := (\theta_n)_{n\in\NN_0}\subset \mathcal{H}$, the sequence $\Xi:= (\xi_n)_{n\in\NN_0}\subset\ell^2(\NN_0)$.
The map $\mathfrak{O}^{\frac12}\!:\! \mathcal{H}\! \to\! \ell^2({\NN}_0)$ is invertible.
 Thus we set $K^{(2)}(\Theta) =:\frac12 Q_{\mbox{\rm{\tiny{E}}}}(\Xi)$, viz.
 \begin{align}\label{eq:QofXI}
\hspace{-1truecm}
Q_{\mbox{\rm{\tiny{E}}}}(\Xi) = &\, {\sum\limits_n} \biggl[
 1 + \lambda\frac{2}{2 n + 1} {\sum\limits_{k=1}^{n}} \frac{\varpi^2}{\varpi^2 + k^2} \biggr] \xi_n^2  \\
\notag &
- \lambda
\sum\!\sum\limits\limits_{\hskip-0.5truecm n\neq m} 
\xi_n  \biggl[\frac{1}{\sqrt{2 n + 1}}\,\frac{\varpi^2} 
{ \varpi^2 + (n-m)^2}\, \frac{1}{\sqrt{2 m + 1}}\biggr] \xi_m \\
\notag &
- \lambda \sum\limits_n\!\sum\limits_m   
\xi_n  \biggl[\frac{1}{\sqrt{2 n + 1}}\, \frac{\varpi^2}{ \varpi^2 +(n+m+1)^2}\, \frac{1}{\sqrt{2 m + 1}}\biggr] \xi_m,
\end{align}
where the contributions from the first line at r.h.s.(\ref{eq:QofXI}) are positive, those from the second and third line
negative.

 We note that $Q_{\mbox{\rm{\tiny{E}}}}(\Xi)$ stated in (\ref{eq:QofXI}) is precisely 
the non-dispersive limit $P(d\Omega) = \delta(\omega-\Omega)d\omega$ of the functional $Q(\Xi)$ of the Eliashberg theory with
dispersive phonons presented in eq.(38) of \cite{KAYdispersive}.
 As such it is endowed with all the characteristics of the functional $Q(\Xi)$ in general that we established in  \cite{KAYdispersive}.
 Namely, given $\lambda>0$ and $\varpi > 0$, the functional $Q_{\mbox{\rm{\tiny{E}}}}$ given in (\ref{eq:QofXI}) has 
a minimum on the sphere $\big\{\Xi\in\ell^2(\NN_0):  \|\Xi\|_{\ell^2}=1\big\}$.
 The minimizing (optimizing) eigenmode $\Xi^{\mbox{\rm\tiny{opt}}}$ satisfies 
$\mathfrak{O}^{-\frac12}\Xi^{\mbox{\rm\tiny{opt}}} \in \RR_+^{\mathbb{N}_0}$. 
  Moreover, for each $\varpi>0$ there is a unique $L_{\mbox{\tiny{E}}}(\varpi)>0$ at which
$\min\big\{Q_{\mbox{\rm{\tiny{E}}}}(\Xi) \!:\!  \|\Xi\|_{\ell^2}=1\big\} =0$, 
and $\min\big\{Q_{\mbox{\rm{\tiny{E}}}}(\Xi) \!:\! \|\Xi\|_{\ell^2}=1\big\} >0$ 
when $\lambda <L_{\mbox{\tiny{E}}}(\varpi)$, while $\min\big\{Q_{\mbox{\rm{\tiny{E}}}}(\Xi) \!:\! \|\Xi\|_{\ell^2}=1\big\} <0$ 
when $\lambda>L_{\mbox{\tiny{E}}}(\varpi)$.
 Furthermore, the map $\varpi\mapsto L_{\mbox{\tiny{E}}}(\varpi)$ is continuous on $\RR_+$.
\smallskip

 The functional $Q_{\mbox{\rm{\tiny{E}}}}(\Xi)$ stated in (\ref{eq:QofXI}) is the quadratic form of a self-adjoint operator.
 Letting $\big\langle \Xi,\widetilde\Xi\big\rangle$ denote the usual $\ell^2(\NN_0)$ inner product between
two $\ell^2$ sequences $\Xi$ and $\widetilde\Xi$, we write $Q_{\mbox{\rm{\tiny{E}}}}$ shorter thus:
 \begin{equation}\label{eq:QofXIrew}
Q_{\mbox{\rm{\tiny{E}}}}(\Xi) =  \bigl\langle \Xi\, , \big( \mathfrak{I} - \lambda \mathfrak{H}\big)  \Xi\bigr\rangle \,.
 \end{equation}
Here, $\mathfrak{I}$ is the identity operator, and $\mathfrak{H} =   - \mathfrak{H}_1 + \mathfrak{H}_2 + \mathfrak{H}_3$, where
the $\mathfrak{H}_j = \mathfrak{H}_j(\varpi)$ for $j\in\{1,2,3\}$ are operators that act as follows, componentwise:
 \begin{align}
\label{eq:HopsCompsONE}
\hspace{-0.8truecm}
(\mathfrak{H}_1(\varpi)\Xi)_n = &\, 
 \biggl[\frac{2}{2 n + 1} {\sum\limits_{k=1}^{n}} \frac{\varpi^2}{\varpi^2 + k^2}\biggr] \xi_n \,, \\
\label{eq:HopsCompsTWO}
\hspace{-0.8truecm}
(\mathfrak{H}_2(\varpi)\Xi)_n = &\, 
\sum\limits_{m\neq n} 
\biggl[\frac{1}{\sqrt{2 n + 1}}\,\frac{\varpi^2} 
{\varpi^2+(n-m)^2}\,\frac{1}{\sqrt{2 m+1}}\biggr]\xi_m\,,\\
\label{eq:HopsCompsTHREE}
\hspace{-0.8truecm}
(\mathfrak{H}_3(\varpi)\Xi)_n = &\, 
\sum\limits_m \biggl[\frac{1}{\sqrt{2 n + 1}}\, \frac{\varpi^2}{ \varpi^2 +(n+m+1)^2}\, \frac{1}{\sqrt{2 m + 1}}\biggr]\xi_m\,.
\end{align}
 Note that $\mathfrak{H}_1$ is a diagonal operator with non-negative diagonal elements, 
$\mathfrak{H}_2$ is a real symmetric operator with vanishing diagonal elements and positive off-diagonal elements,
and $\mathfrak{H}_3$ is a real symmetric operator with all positive elements.

 The operators $\mathfrak{H}_j(\varpi)$, $j\in\{1,2,3\}$, are the dispersionless limit where 
$P(d\omega)=\delta(\omega-\Omega)d\omega$
of the operators $\mathfrak{K}_j(P,T)$, $j\in\{1,2,3\}$ introduced in \cite{KAYdispersive}.
 As such they enjoy the same characteristics as all the operators $\mathfrak{K}_j(P,T)$, $j\in\{1,2,3\}$. 
 Thus, in \cite{KAYdispersive} we established that for $j\in\{1,2,3\}$, each $\mathfrak{H}_j\in \ell^2(\NN_0\times\NN_0)$ for all $\varpi > 0$.
 This means that the operators $\mathfrak{H}_j = \mathfrak{H}_j(\varpi)$ for $j\in\{1,2,3\}$ are Hilbert--Schmidt 
operators that map $\ell^2(\NN_0)$ compactly into $\ell^2(\NN_0)$.
\smallskip

 Specializing our pertinent discusion in \cite{KAYdispersive} to the non-dispersive limit, 
we note that $\min\big\{Q_{\mbox{\rm{\tiny{E}}}}(\Xi) \!:\! \|\Xi\|_{\ell^2}=1\big\} = 0$ iff $\lambda\mathfrak{h}(\varpi)=1$,
with $\mathfrak{h}(\varpi)>0$ denoting the largest eigenvalue of $\mathfrak{H}(\varpi)$.
 Precisely when $\lambda = L_{\mbox{\tiny{E}}}(\varpi)$, with
 \begin{equation}\label{eq:LambdaVPh}
L_{\mbox{\tiny{E}}}(\varpi) =  \tfrac{1}{\mathfrak{h}(\varpi)},
\end{equation}
then the pertinent eigenvalue problem for the minimizing mode $\Xi^{\mbox{\tiny{opt}}}$ of $Q_{\mbox{\rm{\tiny{E}}}}(\Xi)$ reads
\begin{equation}\label{eq:EVeqXiOPT}
\big( \mathfrak{I} -L_{\mbox{\tiny{E}}}(\varpi)\mathfrak{H}\big) \Xi^{\mbox{\tiny{opt}}} = 0,
\end{equation}
which, since $\mathfrak{h} = 1/L_{\mbox{\tiny{E}}}(\varpi)$, is equivalent to 
\begin{equation}\label{eq:EVfixPTeqXiOPT}
\mathfrak{C}\big(\mathfrak{h}(\varpi)\big)\Xi^{\mbox{\tiny{opt}}} = \Xi^{\mbox{\tiny{opt}}}
\end{equation}
where here
\begin{equation}\label{eq:Cop}
\mathfrak{C}\big(\eta\big) := 
\big(\eta\mathfrak{I} +\mathfrak{H}_1\big)^{-1} \big(\mathfrak{H}_2+ \mathfrak{H}_3\big).
\end{equation}
 As in the proof of Theorem~1 in \cite{KAYgamma} one shows that $\mathfrak{C}(\eta)$ for $\eta>0$ is
a compact operator that maps the positive cone $\ell^2_{\geq 0}(\NN_0)$ into itself, in fact mapping any non-zero element of
$\ell^2_{\geq 0}(\NN_0)$ into the interior of $\ell^2_{\geq 0}(\NN_0)$, and that
the spectral radius of $\mathfrak{C}\big(\mathfrak{h}\big)$ equals 1.
 Thus the Krein--Rutman theorem applies and guarantees that the nontrivial 
solution $\Xi^{\mbox{\tiny{opt}}}$ of (\ref{eq:EVfixPTeqXiOPT}) is in the 
positive cone $\ell^2_{\geq 0}(\NN_0)$ (after at most choosing the overall sign), 
hence a perturbation  of the normal state $\underline{\Xi}$ toward the superconducting region.

  This establishes Theorems~$1$ and 2.

 One last useful fact about the spectrum of the operator $\mathfrak{H}(\varpi)$ is the following

\smallskip
\noindent
{\bf Proposition~2}: 
\textsl{Let $\varpi> 0$ be given. Then the largest eigenvalue $\mathfrak{h}(\varpi)$ of
$\mathfrak{H}(\varpi)$ is also the spectral radius $\rho\big(\mathfrak{H}(\varpi)\big)$.}
\smallskip

 The proof of Proposition~2 is implied by the proof of the analogous statement about $\mathfrak{k}(P,T)$ in \cite{KAYdispersive}.

 Proposition~2 allows us to characterize $L_{\mbox{\tiny{E}}}(\varpi)$ as follows:
\begin{equation}\label{eq:LambdaVPrho}
L_{\mbox{\tiny{E}}}(\varpi) =  \tfrac{1}{\rho\big(\mathfrak{H}(\varpi)\big)}.
\end{equation}
 Each of (\ref{eq:LambdaVPrho}) and (\ref{eq:LambdaVPh}) offer their own advantages to estimate $L_{\mbox{\tiny{E}}}$.

\subsection{The upper bounds on~$L_{\mbox{\tiny{E}}}(\varpi)$}\label{sec:upperLambdaB}

\subsubsection{\hspace{-10pt}The upper bounds $L_{\mbox{\tiny{E}}}^{(N)}(\varpi)$ for $N\in\{1,2,3,4\}$}

 We now turn to Theorem~3.
 We only need to specialize the pertinent discusion from \cite{KAYdispersive} to the non-dispersive limit.

 Thus, the variational principle $L_{\mbox{\tiny{E}}}(\varpi) = \frac{1}{\mathfrak{h} (\varpi)}$,
with $\mathfrak{h} (\varpi)>0$ the largest eigenvalue of $\mathfrak{H} (\varpi)$, reads more explicitly as follows:
\begin{equation}\label{eq:LAMBDAcVP}
L_{\mbox{\tiny{E}}}(\varpi) : = \frac{1}{\max_\Xi \frac{\big\langle\Xi\,,\,\mathfrak{H}(\varpi) \,\Xi\big\rangle}
        {\big\langle\Xi,\Xi\big\rangle}},
\end{equation}
where the maximum is taken over non-vanishing $\Xi\in\ell^2(\NN_0)$.

 Since $\mathfrak{H} (\varpi)$ is compact, in principle one can get
arbitrarily accurate upper approximations to $L_{\mbox{\tiny{E}}}(\varpi)$ by restricting 
$\mathfrak{H} (\varpi)$ to suitably chosen finite-dimensional subspaces of $\ell^2(\NN_0)$.
 A sequence of decreasing rigorous upper bounds on $L_{\mbox{\tiny{E}}}(\varpi)$ that converges to $L_{\mbox{\tiny{E}}}(\varpi)$ is
 obtained by restricting the variational principle to a sequence of subspaces of $\ell^2(\NN_0)$ 
of vectors of the type $\Xi_N:= (\xi_0,\xi_1,\dots,\xi_{N-1},0,0,\dots)$, with $\xi_j>0$ for $j\in\{0,...,N-1\}$ and $N\in\NN$.
 Evaluating (\ref{eq:LAMBDAcVP}) with $\Xi_N$ in place of $\Xi^{\mbox{\tiny{opt}}}$ yields a strictly monotonically decreasing
sequence of upper bounds $L_{\mbox{\tiny{E}}}^{(N)}(\varpi)$ on $L_{\mbox{\tiny{E}}}(\varpi)$, viz.
\begin{equation}\label{eq:LAMBDAcVPtrialN}
L_{\mbox{\tiny{E}}}^{(N)}(\varpi) : = 
\frac{1}{\max_{\Xi_N} \frac{\big\langle\Xi_N,\,\mathfrak{H}(\varpi)\,\Xi_N\big\rangle}
        {\big\langle\Xi_N,\,\Xi_N\big\rangle}}.
\end{equation}

 The evaluation of (\ref{eq:LAMBDAcVPtrialN}) is equivalent to finding the largest eigenvalue of
a real symmetric matrix $N\times N$ matrix $\mathfrak{M}$, i.e. the largest zero of the associated
degree-$N$ characteristic polynomial of $\mathfrak{M}$.
 As noted in \cite{KAYgamma}, the coefficients $c_k$ of the characteristic polynomial 
$\det\big(\mu\mathcal{I}-\mathfrak{M}\big) =: \sum_{k=0}^N c_k\mu^k$ 
are explicitly known polynomials of degree $N-k$ in $\mathrm{tr}\, \mathfrak{M}^j$, $j\in\{1,...,N\}$.
 When $N\in\{1,2,3,4\}$ the zeros of the characteristic polynomial can be computed algebraically in closed form.
 For general real symmetric $N\times N$ matrices $\mathfrak{M}$ these spectral formulas have been listed in
\cite{KAYgamma} and need not be repeated here. 
 
 The task that remains is to substitute $\mathfrak{H}^{(N)}$, $N\in\{1,2,3,4\}$, for $\mathfrak{M}$ and to
select the largest eigenvalue for each $N$ from these spectra.
 Since this was done in \cite{KAYdispersive} for the pertinent operators $\mathfrak{K}^{(N)}$, $N\in\{1,2,3,4\}$, 
all that needs to be done is to take the limit where $P(d\omega)=\delta(\omega-\Omega)d\omega$. 
 This yields the formulas for $\mathfrak{h}^{(N)}(\varpi)$, $N\in\{1,2,3,4\}$,
stated in Theorem~3.

 Approximations with $N>4$ require a numerical approximation for each value of $\varpi$ that is of interest.

\subsubsection{The upper bounds $L_{\mbox{\tiny{E}}}^{(N)}(\varpi)$ at $\varpi\!\gg\! 1$ for $N\!\in\!\NN$}

 We here prove Theorem~6 by evaluating $L_{\mbox{\tiny{E}}}{(N)}(\varpi)$ asymptotically, when $\varpi\sim\infty$,
up to the first two significant terms, for all $N\in\NN$.
\smallskip
 
\noindent
\textsl{Proof}: We begin by recalling a proposition of \cite{KAYdispersive}, specialized for the Holstein model.

\smallskip
\noindent
{\bf Proposition~3}: 
\textsl{When $\varpi\to\infty$, we have}
\begin{equation}\label{eq:LambdaNasympINF}
\mathfrak{h}^{(N)}(\varpi) \to \mathfrak{h}^{(N)}(\infty) =  -1+2{\textstyle\sum\limits_{n=0}^{N-1}\frac{1}{2n+1} }.
\end{equation}
\smallskip

 We note that r.h.s.(\ref{eq:LambdaNasympINF}) diverges to $\infty$ when $N\to\infty$, essentially like $\ln N$. 
 Thus, $L_{\mbox{\tiny{E}}}^{(N)}(\infty)=\frac{1}{\mathfrak{h}^{(N)}(\varpi)} \to 0$ as $N\to \infty$, as claimed in the introduction.
\smallskip

 The next proposition is novel.
\smallskip

\noindent
{\bf Proposition~4}: 
\textsl{In the limit when $\varpi\to\infty$, we have}
\begin{equation}\label{eq:LambdaNasympINFtwo}
\Big(\mathfrak{h}^{(N)}(\varpi) - \mathfrak{h}^{(N)}(\infty)\Big) \varpi^2
\to
B_N 
\end{equation}
with $B_N$ given in (\ref{eq:BN}).
\smallskip

\noindent
\textsl{Proof of Proposition-4}: By Taylor series expansion of $\mathfrak{H}^{(N)}(\varpi)$ in powers of $1/\varpi^2$ about 
$1/\varpi^2=0$, one obtains
\begin{equation}\label{eq:HNasympINFinf}
\lim_{\varpi\to\infty} \varpi^2\Big(\mathfrak{H}^{(N)}(\varpi) + \mathfrak{I}^{(N)} - 2\, \Xi_N^*\otimes\Xi_N^*\Big)
= \mathfrak{L}^{(N)}.
\end{equation}
 By first-order perturbation theory \cite{Kato}, 
\begin{equation}\label{eq:hNasympINFinf}
\lim_{\varpi\to\infty} \varpi^2 \Big(\mathfrak{h}^{(N)}(\varpi) - \mathfrak{h}^{(N)}(\infty)\Big) 
= \frac{\left\langle \Xi_N^*, \mathfrak{L}^{(N)} \Xi_N^*\right\rangle}{\left\langle \Xi_N^*, \Xi_N^*\right\rangle},
\end{equation}
with $\mathfrak{L}^{(N)} := \mathfrak{L}^{(N)}_1 - \mathfrak{L}^{(N)}_2 - \mathfrak{L}^{(N)}_3$
acting componentwise as follows:
 \begin{align}
\label{eq:LopONE}
\hspace{-0.8truecm}
(\mathfrak{L}^{(N)}_1\Xi^*)_n = &\, 
 \biggl[\frac{2}{2 n + 1} {\sum\limits_{k=1}^{n}} k^2\biggr] \xi_n^* \,, \\
\label{eq:LopTWO}
\hspace{-0.8truecm}
(\mathfrak{L}^{(N)}_2\Xi^*)_n = &\, 
\sum\limits_{m\neq n} 
\biggl[\frac{(n-m)^2}{\sqrt{2 n + 1}\,\sqrt{2 m+1}}\biggr]\xi_m^*\,,\\
\label{eq:LopTHREE}
\hspace{-0.8truecm}
(\mathfrak{L}^{(N)}_3\Xi^*)_n = &\, 
\sum\limits_m \biggl[\frac{(n+m+1)^2}{\sqrt{2 n + 1}\, \sqrt{2 m + 1}}\biggr]\xi_m^*\,.
\end{align}

 Evaluation of (\ref{eq:hNasympINFinf}) yields (\ref{eq:LambdaNasympINFtwo}) with $B_N$ given in (\ref{eq:BN}).
 \hfill {\textbf{Q.E.D.}}

Propositions~3 and~4 prove Theorem~6.  \hfill {\textbf{Q.E.D.}}
\smallskip

\subsubsection{The upper bounds $L_{\mbox{\tiny{E}}}^{(N)}(\varpi)$ at $\varpi\!\ll\! 1$ for $N\!\in\!\NN$}

We now turn to Theorem~7.

 The proof of  Theorem~7 is contained in the proof of Theorem~7 of \cite{KAYdispersive}, which
yields the small $\varpi$ expansion
\begin{equation}
\mathfrak{H}_j(\varpi)= \varpi^2 \mathfrak{G}_j(2) - \varpi^4\mathfrak{G}_j(4) \pm \cdots,\quad j\in\{1,2,3\},
\end{equation}
and the analogous expansion for their $N$-frequency truncations, then applies first-order perturbation theory \cite{Kato},
and finally establishes that for all $N\in\NN$ we {have $\langle \mathfrak{G}^{(N)}(4)\rangle^{}_{\!2}>0$,} where
\begin{equation}\label{eq:expectKfourWITHtwo}
\langle \mathfrak{G}^{(N)}(4)\rangle^{}_{\!2}
 := \frac{\big\langle\Xi^{\mbox{\tiny{opt}}}_N(2),\mathfrak{G}^{(N)}(4)\,\Xi^{\mbox{\tiny{opt}}}_N(2)\big\rangle}
        {\big\langle\Xi^{\mbox{\tiny{opt}}}_N(2),\Xi^{\mbox{\tiny{opt}}}_N(2)\big\rangle} ;
\end{equation}
here, $\Xi^{\mbox{\tiny{opt}}}_N(2)$ denotes the eigenvector for the maximal eigenvalue $\mathfrak{g}^{(N)}(2)$ 
of $\mathfrak{G}^{(N)}(2)$. 
 The inequality {$\langle \mathfrak{G}^{(N)}(4)\rangle^{}_{\!2}>0$}
is a consequence of the following stronger result proved in \cite{KAYdispersive}.
\smallskip

\noindent
{\bf Proposition~5}: \textsl{Let $\gamma > 0$ be given. Then for all $\gamma^\prime >0$ and $N\in\NN_0$,}
\begin{equation}\label{eq:expectKgammaWITHgammaNULL}
\langle \mathfrak{G}^{(N)}(\gamma^\prime)\rangle^{}_{\!\gamma}
:= \frac{\big\langle\Xi^{\mbox{\tiny{opt}}}_N(\gamma),\mathfrak{G}^{(N)}(\gamma^\prime)\,\Xi^{\mbox{\tiny{opt}}}_N(\gamma)\big\rangle}
        {\big\langle\Xi^{\mbox{\tiny{opt}}}_N(\gamma),\Xi^{\mbox{\tiny{opt}}}_N(\gamma)\big\rangle} {> 0},
\end{equation}
\textsl{with}
$\Xi^{\mbox{\tiny{opt}}}_N(\gamma)$ \textsl{any eigenvector of the
top eigenvalue $\mathfrak{g}^{(N)}(\gamma)$ of $\mathfrak{G}^{(N)}(\gamma)$.}
\smallskip

This establishes Theorem~7.

\subsection{The rigorous lower bound on~$L_{\mbox{\tiny{E}}}(\varpi)$}\label{sec:lowerLambdaB}

 Turning to Theorem~4, it suffices to note that 
its proof is included as limiting case $P(d\omega)\to \delta(\omega-\Omega)d\omega$ with $\Omega>0$ of
the proof of Theorem~6 in \cite{KAYdispersive}. 
 
 We add the remark that our lower bound $L_{\mbox{\tiny{E}}}^*(\varpi)$
on $L_{\mbox{\tiny{E}}}(\varpi)$ is a uniform lower bound on the analogous function in standard Eliashberg theory
with dispersive phonons, i.e. on $\Lambda(P,T)$, for all $P(d\omega)\in\mathcal{P}$ 
supported on $[0,\Omega]$.
 This follows by inspection of the proof of Theorem~6 and its corollaries in \cite{KAYdispersive}.

\subsection{From $L_{\mbox{\tiny{E}}}(\varpi)$ to $T_c^{}(\lambda,\Omega)$}\label{sec:TcA}

 We turn to Theorem~5. 

 The proof of Theorem~5 is largely a special case of the pertinent proofs of Proposition~1, Theorem~2, and Corollary~1 
in \cite{KAYdispersive}. 
 Indeed, the monotonicity of $\varpi\mapsto L_{\mbox{\tiny{E}}}(\varpi)$ for $\varpi\in[0,\sqrt{2}]$ follows simply by the
specification $P(d\omega)=\delta(\omega-\Omega)d\omega$ 
in the proof of monotonicity of $T\mapsto\Lambda(P,T)$ for $T>T_*(P)$, given $P$, in \cite{KAYdispersive}.
 Also the bound $\lambda_* \leq 0.767...$ stated in Theorem~5, i.e. of
the upper estimate of the left boundary of the interval of $\lambda$ values for which a unique critical temperature $T_c(\lambda,\Omega)$
is guaranteed by the monotonicity of $\varpi\mapsto L_{\mbox{\tiny{E}}}(\varpi)$ for when $\varpi\in[0,\sqrt{2}]$, is obtained simply
by evaluation of the bound stated in Corollary~1 of \cite{KAYdispersive} with $P(d\omega)=\delta(\omega-\Omega)d\omega$,
followed by decimal expansion.
 One also needs to note that with $P(d\omega)=\delta(\omega-\Omega)d\omega$ the upper estimate of $T_*(P)$ in Theorem~2 of \cite{KAYdispersive} 
becomes $\Omega/2\sqrt{2}\pi$.

\subsection{Lower bounds on~$T_c(\lambda,\Omega)$}

 We here get to the part of Corollary~1 that follows from Theorem~4.
 The validity of Corollary~1 is largely obvious, so we confine ourselves to some additional remarks.

\subsubsection{The lower bound $T_c^{(1)}(\lambda,\Omega)$}

 The lower $T_c$ bound (\ref{eq:TcONE}) follows easily from the lower bound (\ref{eq:hONE}) on $\mathfrak{h}(\varpi)$
and the characterization of $L_{\mbox{\tiny{E}}}(\varpi)$ as reciprocal value of $\mathfrak{h}(\varpi)$.
 Indeed, the map $\varpi\mapsto$r.h.s.(\ref{eq:hONE}) is obviously monotone increasing, hence invertible for all $\varpi>0$.
 It is readily inverted and yields (\ref{eq:TcONE}), restricted to $\lambda>1$.
 This bound was previously obtained in \cite{AllenDynes}, 
by discussing a truncation to a single Matsubara frequency of the linearized Eliashberg gap equations in their
original model formulation.

\subsubsection{The lower bound $T_c^{(2)}(\lambda,\Omega)$}\label{sec:TcTWO}

 The formula (\ref{eq:TcLOWERboundsNoneTOfour}) for $N=2$ does not have a closed form expression 
in terms of algebraic functions, as we will see.
 As far as we can tell, it does not seem to have a closed form expression in known special functions, either.
 Yet its parameter representation (\ref{eq:tildeCcritPARAM}) is readily discussed. 

 For the $2\times2$ matrix given by the upper leftmost $2\times2$ block of r.h.s.(\ref{eq:Hfour})
the largest eigenvalue (\ref{eq:hTWO}) can be written explicitly as function of $\varpi$ 
with the help of the formulas for the invariants ${\rm tr}\,\mathfrak{H}^{(2)}(\varpi)$ and 
$\det\mathfrak{H}^{(2)}(\varpi)$ listed in Appendix A.1.
  Recalling the abbreviations ${[\![}n{]\!]}(\varpi):= \frac{\varpi^2}{n^2 +\varpi^2}$ for $n\in\NN$, after some algebraic
manipulations we find
\begin{equation}\label{eq:traceHtwo}
{\rm tr}\,\mathfrak{H}^{(2)} = \frac13 \big({[\![} 1{]\!]} + {[\![} 3{]\!]}\big)
\end{equation}
and
\begin{equation}\label{eq:determinantHtwo}
\det\mathfrak{H}^{(2)} = - \frac13 
\left(\!\big({[\![} 1{]\!]} + {[\![} 2{]\!]}\big)^2\! 
+ {[\![} 1{]\!]}
 \!\left(2  {[\![} 1{]\!]} - {[\![} 3{]\!]} \right)\! \right).
\end{equation}
 Note that (\ref{eq:traceHtwo}) reveals that ${\rm tr}\,\mathfrak{H}^{(2)}>0$;
note furthermore that $n\mapsto \frac{\varpi^2}{n^2 +\varpi^2}>0$ is strictly decreasing with increasing $n\in\NN$,
and so (\ref{eq:determinantHtwo}) reveals that $\det\mathfrak{H}^{(2)}<0$. 
 Inserting  (\ref{eq:traceHtwo}) and (\ref{eq:determinantHtwo}) into the formula
\begin{equation}\label{eq:hTWOexplicit}
\mathfrak{h}^{(2)}(\varpi) = 
\tfrac12\Big(
{\rm tr}\,\mathfrak{H}^{(2)} + \sqrt{\big({\rm tr}\,\mathfrak{H}^{(2)}\big)^2 - 4 \det\mathfrak{H}^{(2)}}\,\Big),
\end{equation}
then taking its reciprocal, yields the upper bound $L_{\mbox{\tiny{E}}}^{(2)}(\varpi)$ on $L_{\mbox{\tiny{E}}}(\varpi)$ explicitly
\begin{align}\label{eq:LambdaTWOexpl}
 & \hspace{-1truecm} L_{\mbox{\tiny{E}}}^{(2)} =  \\   \notag & \hspace{-0.7truecm}
\frac{6}{{[\![} 1{]\!]} + {[\![} 3{]\!]}  + \sqrt{\big({[\![} 1{]\!]} + {[\![} 3{]\!]}\big)^2\! + 
 12\left(\! \big({[\![} 1{]\!]} + {[\![} 2{]\!]}\big)^2\!
+ {[\![} 1{]\!]}\!\left(2  {[\![} 1{]\!]} - {[\![} 3{]\!]} \right)\! \right)}}\,.
\end{align}
 The map $\varpi\mapsto L_{\mbox{\tiny{E}}}^{(2)}(\varpi)$ is readily seen to be continuous.
 The fact that it also is strictly decreasing when $\varpi> 0$ increases from 0 to $\infty$ is a special non-dispersive limit
case of the analogous monotonicity result proved in \cite{KAYdispersive} for the Eliashberg model with dispersive phonons.

 Therefore, as $\varpi$ runs from $0$ to $\infty$, the map $\varpi \mapsto$r.h.s.(\ref{eq:LambdaTWOexpl})
is continuous and strictly monotonically decreasing to $\lambda_2$, given by (\ref{eq:lambdaN}) for $N=2$; viz.
\begin{align}\label{eq:LambdaTWOexplEVALinfty}
 \lambda_2 =  \tfrac{3}{5}= 0.6\,. 
\end{align}
 
 It follows that the map $\varpi\mapsto\lambda=L_{\mbox{\tiny{E}}}^{(2)}(\varpi)$ is invertible, 
and recalling that $\varpi=\frac{\Omega}{2\pi T}$, this yields a unique lower critical-temperature 
bound $T_c^{(2)}(\lambda,\Omega)$ which is directly propotional to $\Omega$ and increasing in $\lambda$ on its
domain of definition $[\lambda_2,\infty)$.

 We remark that the inversion of $\varpi\mapsto\lambda=L_{\mbox{\tiny{E}}}^{(2)}(\varpi)$ is equivalent to finding
a particular root to a polynomial in $\varpi^2$ of degree $\gg 4$, which is known not to be expressible in 
closed form algebraically.
 
\subsubsection{The lower bound $T_c^{(3)}(\lambda,\Omega)$}

 For the $N=3$ frequencies approximation we have not found a way to write the $\varpi$ dependence of $\mathfrak{h}^{(3)}(\varpi)$
explicitly in a manner that is more condensed than the formulas given in (\ref{eq:tildeCcritPARAM}) for $N=3$,
 supplemented by the formulas of Appendix A.2 for the invariants of the $3\times3$ matrix $\mathfrak{H}^{(3)}(\varpi)$.
 All the same, by our Theorem~4 we know that for $\varpi\leq \sqrt{2}$ the map $\varpi\mapsto\mathfrak{h}^{(3)}(\varpi)$
is strictly monotonic increasing, and by reasoning analogously to how we argued in the paragraph before Corollary~1, 
we conclude that for $\varpi\leq\sqrt{2}$ the map $\varpi\mapsto \mathfrak{h}^{(3)}(\varpi)$ 
is invertible to yield for $\lambda> 1/\mathfrak{h}^{(3)}(\sqrt{2})$ a $T_c^{(3)}(\lambda,\Omega)$ that is 
proportional to $\Omega$ and strictly monotonic increasing in $\lambda$.
 Evaluation yields $1/\mathfrak{h}^{(3)}(\sqrt{2}) = 1.0158...$. 
 Moreover, we know by Theorem~6 that in a small right neighborhood of $\lambda_3 \equiv \frac{15}{31}=0.48387...$ 
our explicit parameter representation for $\tilde{\mathscr{C}}_c^{(3)}$ yields a $T_c^{(3)}(\lambda,\Omega)$ 
that is proportional to $\Omega$ and strictly monotonic increasing in $\lambda$.
 With some extra (not too hard, but daunting) work, one should be able to rigorously prove that 
$\lambda\mapsto T_c^{(3)}(\lambda,\Omega)$ is strictly monotonic increasing for all $\lambda\geq \lambda_3$, 
but here we are content with pointing out that the plot of 
our parameter representation for $T_c^{(3)}(\lambda,\Omega)$ in Fig.~1 reveals that there is no sudden 
``horizontal oscillation'' in the critical curve $\tilde{\mathscr{C}}_c^{(3)}$ for $\lambda\in (0.4838...,1.0158...)$.
\vspace{-10pt}

\subsubsection{The lower bound $T_c^{(4)}(\lambda,\Omega)$}
\vspace{-5pt}

 Essentially everything we wrote about the lower bound $T_c^{(3)}(\lambda,\Omega)$ carries over to the 
lower bound $T_c^{(4)}(\lambda,\Omega)$, by analogy.
 Minor adjustments compared to the $N=3$ approximation
are that $T_c^{(4)}(\lambda,\Omega)$ is well defined for $\lambda> 1/\mathfrak{h}^{(4)}(\sqrt{2}) = 0.7670...$,
while $\lambda_4 =\frac{105}{247}=0.4251...$, and
that the plot of our parameter representation for $T_c^{(4)}(\lambda,\Omega)$ in Fig.~1 reveals that there 
is no sudden  ``horizontal oscillation'' in the critical curve $\tilde{\mathscr{C}}_c^{(4)}$ for 
$\lambda\in (0.4251...,0.7670...)$.
\vspace{-10pt}

\subsection{Upper bounds on $T_c(\lambda,\Omega)$}

\subsubsection{The upper bound $T_c^*(\lambda,\Omega)$}

Our proof of Theorem~5 extablishes rigorously an explicit lower bound $1/\mathfrak{h}^*(\varpi)$
on $L_{\mbox{\tiny{E}}}(\varpi)$, with $\mathfrak{h}^*(\varpi)$ given in (\ref{eq:hSTAR}).
 Since $\varpi\mapsto \mathfrak{h}^*(\varpi)$ is manifestly strictly monotonically increasing with $\varpi$,
this map is invertible, moreover explicitly so in closed form.
 This yields the upper bound on $T_c$ given as function of $\lambda$ and $\Omega$ in eq.(\ref{eq:TcSTAR})
of Corollary~1.
\newpage

 For large $\lambda\sim\infty$ this bound is $\sim C\sqrt{\lambda}$ with a $C$ that is larger than the optimal coefficient
in Conjecture~1 by a factor $\approx 2.034$.

\subsubsection{The large-$\lambda$ upper bound $T_c^\sim(\lambda,\Omega)$}
 
The discussion in section 2 establishes that $T_c^\sim(\lambda,\Omega)$ in Conjecture~1
is an upper bound on $T_c(\lambda,\Omega)$ for large enough $\lambda$. 
 This is a consequence of Corollary~3 to Theorem~7, which establishes Proposition~1.

 Recall that Conjecture~1 proposes that $T_c^\sim(\lambda,\Omega)$ is an upper bound on $T_c(\lambda,\Omega)$
for all $\lambda>0$ and $\Omega>0$. 
 Fig.~1 and Fig.~2 present numerical evidence for its veracity.
\vspace{-10pt}

\section{Summary and Outlook}\label{sec:sumANDout}
\vspace{-5pt}

\subsection{Summary}

 In this paper we rigorously studied the phase transition between {normal and superconducting states}
in a representative version of the standard Eliashberg theory in which
the effective electron-electron interactions are mediated 
by {dispersion-free} Einstein phonons of frequency $\Omega>0$, having electron-phonon coupling strength $\lambda>0$.
 The model is obtained by taking the dispersionless limit of the standard Eliashberg model in which 
the effective electron-electron interactions are mediated 
by phonons with {Eliashberg spectral function} $\alpha^2\!F(\omega)$ that defines the electron-phonon coupling strength $\lambda>0$.
 The standard Eliashberg model we studied in \cite{KAYdispersive}.
 The results obtained in the present paper are mostly special cases of the results of \cite{KAYdispersive}.
 We emphasize that our results for the Eliashberg model with Einstein phonons 
are more detailed and quantitative than those of \cite{KAYdispersive}, which 
remained rather qualitative since  $\alpha^2\!F(\omega)$ was left largely unspecified. 

 After a suitable rescaling with $\lambda$ the Eliashberg model with Einstein phonons 
is asymptotic to the $\gamma$ model at $\gamma=2$ when $\lambda\to\infty$.
 The $\gamma$ model was studied in our previous paper \cite{KAYgamma}.

 We {showed in this paper } that the normal and the superconducting regions in
the positive $(\lambda,\Omega,T)$ octant are both simply connected, and separated 
by a critical surface $\mathscr{S}_c$ that is a ruled graph over the positive $(\Omega,T)$ quadrant. 
 It is given by a function $\lambda = \Lambda_{\mbox{\tiny{E}}}(\Omega,T)$ that depends on $\Omega$ and $T$ only through the 
combination $\Omega/ T =: 2\pi\varpi$, thus $\Lambda_{\mbox{\tiny{E}}}(\Omega,T)=L_{\mbox{\tiny{E}}}(\varpi)$.
 Therefore, the critical surface $\mathscr{S}_c$ is completely characterized by a critical curve
${\mathscr{C}}_c$ in the positive $(\lambda,\varpi)$ quadrant that is a graph over the $\varpi$ axis, viz.
${\mathscr{C}}_c = \{(\lambda,\varpi)\in\RR_+^2: \lambda = L_{\mbox{\tiny{E}}}(\varpi)\}$.

 We furthermore showed that $L_{\mbox{\tiny{E}}}(\varpi) = 1/\mathfrak{h}(\varpi)$, where $\mathfrak{h}(\varpi)>0$ is the largest
eigenvalue of an explicitly constructed compact operator $\mathfrak{H}(\varpi)$ on $\ell^2(\NN_0)$, where $\NN_0$ is 
the set of non-negative integers that enumerates the positive Matsubara frequencies.
 Since a compact operator on a separable Hilbert space can be arbitrarily closely approximated by truncating it to finite-dimensional
subspaces, in this case spanned by the first $N$ positive Matsubara frequencies, we obtained 
from our variational principle a strictly monotonically decreasing sequence of rigorous upper bounds on $L_{\mbox{\tiny{E}}}(\varpi)$,
the first four of which we have computed explicitly in closed form.

 Through spectral estimates of $\mathfrak{h}(\varpi)$ from above we also rigorously obtained an explicit 
lower bound on $L_{\mbox{\tiny{E}}}(\varpi)$.
 
 Physical intuition, based on empirical evidence, suggests that the phase transition can be characterized in terms of a 
\textsl{critical temperature} $T_c(\lambda,\Omega)$, which is equivalent to saying the critical surface $\mathscr{S}_c$ 
is a graph over the positive $(\lambda,\Omega)$ quadrant.
 This in turn is equivalent to saying that the map $\varpi\mapsto L_{\mbox{\tiny{E}}}(\varpi)$ is strictly monotone, hence invertible to yield
$\varpi= L_{\mbox{\tiny{E}}}^{-1}(\lambda)$.
 Recalling the definition of $\varpi$, this would give $T_c(\lambda,\Omega) = \Omega f(\lambda)$ with
$f(\lambda)= \frac{1}{2\pi L_{\mbox{\tiny{E}}}^{-1}(\lambda)}$.
 By taking the dispersionfree limit of our results in \cite{KAYdispersive}, we showed that all our upper approximations to 
the map $\varpi\mapsto L_{\mbox{\tiny{E}}}(\varpi)$, and this map itself, 
are strictly monotone decreasing for $\varpi\in[0,\sqrt{2}]$.
 We also supplied compelling evidence for the conjecture that the map $\varpi\mapsto L_{\mbox{\tiny{E}}}(\varpi)$
is strictly monotone decreasing for all $\varpi\in[0,\infty)$, but to prove it would require a different strategy.

 Since the explict fourth upper bound on $L_{\mbox{\tiny{E}}}(\varpi)$ yields $L_{\mbox{\tiny{E}}}^{(4)}(\sqrt{2})= 0.7670...$, 
what we just wrote proves that a unique critical temperature $T_c(\lambda,\Omega)$ in the Eliashberg model with Einstein phonons is 
mathematically well-defined in terms of the untruncated linearized Eliashberg gap equations whenever $\lambda>0.7670...$.
 Also this $\lambda$ value is not a sharp boundary but a consequence of our method of proof.
 While mathematically desirable to prove the existence of a unique $T_c(\lambda,\Omega)$ for all
$\lambda>0$, from a theoretical physics perspective the range $\lambda>0.7670...$ covers all cases of interest so far.
 Moreover, as detailed above already, on the interval $\lambda>0.7670...$ the critical temperature $T_c(\lambda,\Omega)$
takes the form $T_c(\lambda,\Omega) = \Omega f(\lambda)$, and $f(\lambda) =\frac{1}{2\pi L_{\mbox{\tiny{E}}}^{-1}(\lambda)}$
is strictly monotonic increasing with $\lambda$, asymptotically for large $\lambda$ like $C\sqrt{\lambda}$, with 
$C=\frac{1}{2\pi}\sqrt{\mathfrak{g}(2)}=0.1827262477...$, where $\mathfrak{g}(2)$ is the spectral radius of a 
compact operator $\mathfrak{G}(2)$ associated with the $\gamma$ model for $\gamma=2$. 
{This monotonicity {had been anticipated previously and} widely used in the 
superconductivity literature, {though} without proof. 
In particular, it has been instrumental {in} obtaining bounds on $T_c$ based on the limits of applicability of the 
Eliashberg theory to physical systems \cite{Semenok,EKS,EsterlisETal,CAEK,Sad,Tra,YuzAltPat} 
(in contrast to bounds from \textsl{within} this theory that we constructed here).}

 With the existence of a unique critical temperature secured for most situations of interest, we have the following 
interesting application of our results (pretending the Einstein phonon model would accurately capture the behavior of some
superconductors in the laboratory).
 Namely, measuring the phonon frequency $\Omega$ and the critical temperature $T_c$ yields the electron-phonon coupling
constant through our formula
\begin{equation}\label{eq:lambdaOFcritT}
\lambda = L_{\mbox{\tiny{E}}}(\Omega,T_c),
\end{equation}
with the function $L_{\mbox{\tiny{E}}}$ determined by our variational principle (\ref{eq:lambdaVP}), see Theorem~3.
 Since there has not been any experimental means yet to measure the electron-phonon coupling constant directly, 
our formula (\ref{eq:lambdaOFcritT}) provides a useful algorithm to obtain it from the easy-to-measure quantities $\Omega$ and
$T_c$.

{We conclude with the remark that our assessment, in the summary section in \cite{KAYdispersive} 
of our upper and lower bounds on $T_c$ in the standard Eliashberg theory with generally dispersive phonons, in regard to 
the existing superconductivity literature applies also to the nondispersive limit of Einstein phonons.
 It need not be repeated here.}
\vspace{-5pt}

\subsection{Outlook}

{The present paper completes our series of three papers on the rigorous study of the Eliashberg gap equations as 
linearized about the normal state.
 There are further issues concerning the linearized Eliashberg gap equations that merit clarification or vindication, 
but these fall outside the thrust of our three papers about bounds on $T_c$ for various realizations of Eliashberg theory,
and will be addressed elsewhere.}

{Our next goal is the} study of the non-linear Eliashberg gap equations.
 It should not come as a surprise that the results on the linearized Eliashberg gap equations will play an important role in our
study of the nonlinear equations, too. 
 However, in general our Hilbert space analysis of papers I-III will have to be replaced by an analysis of operators in a certain
 Banach space.

\vfill

\noindent
{\bf Acknowledgement}: We thank the two referees for their constructive criticisms. We also thank Steven Kivelson for his comments. 
\bigskip

DATA AVAILABILITY STATEMENT: No data have been produced for this paper.

CONFLICT OF INTEREST STATEMENT: The authors declare that they have no conflict of interest.

\begin{appendices}

\section{The matrix invariants}\label{appendixA} 

 In this appendix we list the matrix invariants that enter our explicit spectral formulas.

\subsection{Trace and determinant for $\mathfrak{H}^{(2)}$}

With the help of Maple, we computed 
\begin{equation}\label{eq:trHtwo}
{\rm tr}\,\mathfrak{H}^{(2)} =
 \frac23 \frac{\varpi^2(5 + \varpi^2)}{(1+\varpi^2)(3^{2}+\varpi^2)}
\end{equation}
and
\begin{equation}\label{eq:detHtwo}
\det\mathfrak{H}^{(2)} = 
 - \frac13 \frac{\varpi^4( 497 + 357\varpi^2 + 81 \varpi^{4} + 5 \varpi^{6})}
{(1+\varpi^2)^2(2^{2}+\varpi^2)^2(3^{2}+\varpi^2)}.
\end{equation}

\subsection{${\rm tr}\, \mathfrak{H}^{(3)}$, ${\rm tr\, adj}\, \mathfrak{H}^{(3)}$, and $\det \mathfrak{H}^{(3)}$}

 With the help of Maple, we computed 
\begin{align}\label{eq:TR3x3H}
{\rm tr}\,\mathfrak{H}^{(3)} = 
\frac{1}{15}\frac{(4^2+\varpi^2)}{\prod\limits_{j=1}^5(j^2+\varpi^2)}
 \sum_{j=0}^3 P_j \varpi^{2j+2},
\end{align}
with
\begin{align}\label{eq:trHthreePnull}
P_0 &= -1642, \\ 
\label{eq:trHthreePeins}
P_1 & = -1123,\\ 
\label{eq:trHthreePzwei}
P_2 &= -56,\\ 
\label{eq:trHthreePdrei}
P_3 & = 1,
\end{align}
and where
\begin{align}\label{eq:tradjHthree}
 \hspace{-1truecm}
{\rm tr\, adj}\,\mathfrak{H}^{(3)} =  
 \frac{1}{15} 
\frac{(5^{2}+\varpi^2)}{\prod\limits_{j=1}^5 (j^2+\varpi^2)^2} 
 \sum_{j=0}^7 Q_j \varpi^{2j+4},
\end{align}
with
\begin{align}
\label{eq:tradjHthreeQnull}
Q_0 &= - 178415760\\
\label{eq:tradjHthreeQeins}
Q_1 &= -184933048,\\
\label{eq:tradjHthreeQzwei}
\hspace{-10pt}
Q_2 & = -76880761\\
\label{eq:tradjHthreeQdrei}
Q_3 & = -16105091,\\
\label{eq:tradjHthreeQvier}
Q_4 &= - 1840578, \\
\label{eq:tradjHthreeQfuen}
Q_5 &=  - 115414 ,\\
\label{eq:tradjHthreeQsech}
Q_6 &= - 3701\\ 
\label{eq:tradjHthreeQsieb}
Q_7 & = -47,
\end{align}
and where
\begin{align}\label{eq:detHthree}
\hspace{-1truecm} 
\det\mathfrak{H}^{(3)} = 
 \frac{1}{15} \frac{(4^{2}+\varpi^2)(5^{2}+\varpi^2)^2}{\prod\limits_{j=1}^5(j^2+\varpi^2)^3}
\sum_{j=0}^9 R_j \varpi^{2j+6},
\end{align}
with
\begin{align}\label{eq:detHthreeRnull}
R_0 &= 2558100032,\\
\label{eq:detHthreeReins}
R_1 &= 4173421560,\\
\label{eq:detHthreeRzwei}
R_2 &=  2816977328,\\
\label{eq:detHthreeRdrei}
R_3 &= 1019355095 ,\\
\label{eq:detHthreeRvier}
R_4 & = 217124598,\\
\label{eq:detHthreeRfuen}
R_5 &= 28353481,\\
\label{eq:detHthreeRsech}
R_6 &= 2283172,\\
\label{eq:detHthreeRsieb}
R_7 &= 109833,\\
\label{eq:detHthreeRacht}
R_8 &= 2870,\\
\label{eq:detHthreeRneun}
R_9 &= 31.
\end{align}

\subsection{${\rm tr}\, \big(\mathfrak{H}^{(4)}\big)^j$ for $j\in\{1,2,3\}$, and $\det \mathfrak{H}^{(4)}$}

With the help of Maple we computed
\begin{align}\label{eq:TR4x4H}
{\rm tr}\,\mathfrak{H}^{(4)} = 
\frac{4}{105}\frac{(4^2+\varpi^2)(6^2+\varpi^2)}{\prod\limits_{j=1}^7(j^2+\varpi^2)}
\sum_{j=0}^4 A_j \varpi^{2j+2},
\end{align}
with
\begin{align}
\label{eq:trHfourAnull}
A_0 &= - 587614, \\ 
\label{eq:trHhourAeins}
A_1 & = -378887,\\ 
\label{eq:trHfourAzwei}
A_2 &= -48741,\\ 
\label{eq:trHfourAdrei}
A_3 & = -1741, \\
\label{eq:trHfourAvier}
A_4 & = -17, \\
\end{align}
and where
\begin{align}
\label{eq:TR4x4Hsqared}
{\rm tr}\big(\mathfrak{H}^{(4)}\big)^2 = 
\frac{4}{105^2}\frac{1}{\prod\limits_{j=1}^7(j^2+\varpi^2)^2}
\sum_{j=0}^{12} B_j \varpi^{2j+4},
\end{align}
with
\begin{align}
\label{eq:trHfourBnull}
B_0 &=  5528840384999510784, \\ 
\label{eq:trHhourBeins}
B_1 &= 6599696503410581760 , \\ 
\label{eq:trHfourBzwei}
B_2 &= 3326645732455221568,\\ 
\label{eq:trHfourBdrei}
B_3 & = 918435365485009440, \\
\label{eq:trHfourBvier}
B_4 & = 155456935099854829, \\
\label{eq:trHfourBfuenf}
B_5 & = 17126757404561210, \\
\label{eq:trHfourBsechs}
B_6 & = 1269043196251444, \\
\label{eq:trHfourBsieben}
B_7 & = 64176958757030, \\
\label{eq:trHfourBvacht}
B_8 & = 2212609975954, \\
\label{eq:trHfourBneun}
B_9 & = 51013415750, \\
\label{eq:trHfourBzehn}
B_{10} & = 750481900, \\
\label{eq:trHfourBelf}
B_{11} & = 6354810, \\
\label{eq:trHfourBzwoelf}
B_{12} & = 23521, 
\end{align}
and where
\begin{align}\label{eq:TR4x4Hcubed}
{\rm tr}\big(\mathfrak{H}^{(4)}\big)^3 = 
\frac{4}{105^3}\frac{(4^2+\varpi^2)(6^2+\varpi^2)}{\prod\limits_{j=1}^7(j^2+\varpi^2)^3}
\sum_{j=0}^{16} C_j \varpi^{2j+6},
\end{align}
with
\begin{align}\label{eq:trHfouCnull}
C_0 &= 9010878250269017144157696, \\ 
\label{eq:trHhourCeins}
C_1 & = 16624865829373037483712768,\\ 
\label{eq:trHfourCzwei}
C_2 &= 13540239499558750620520256,\\ 
\label{eq:trHfourCdrei}
C_3 & = 6364419134756953975352320, \\
\label{eq:trHfourCvier}
C_4 & = 1926113327143598489598662, \\
\label{eq:trHfourCfuenf}
C_5 & = 398191036579545928245331, \\
\label{eq:trHfourCsechs}
C_6 & = 58319255322911645778671, \\
\label{eq:trHfourCsieben}
C_7 & = 6195524225171893946797, \\
\label{eq:trHfourCvacht}
C_8 & = 484500448290278415303, \\
\label{eq:trHfourCneun}
C_9 & = 28103545935504703422, \\
\label{eq:trHfourCzehn}
C_{10} & = 1210147834157786502, \\
\label{eq:trHfourCelf}
C_{11} & = 38427425846357114, \\
\label{eq:trHfourCzwoelf}
C_{12} & = 885377791146772, \\
\label{eq:trHfourCdreizehn}
C_{13} & = 14357222218367, \\
\label{eq:trHfourCvierzehn}
C_{14} & = 155016577051, \\
\label{eq:trHfourCfuenfzehn}
C_{15} & = 998583881, \\
\label{eq:trHfourCsechzehn}
C_{16} & = 2899087, 
\end{align}
and where
\begin{align}\label{eq:DWR4x4H}
{\rm det} \mathfrak{H}^{(4)} = 
\frac{4}{105}\frac{\prod\limits_{j=5}^7(j^2+\varpi^2)^{j-4}}{\prod\limits_{j=1}^7(j^2+\varpi^2)^4}
 \sum_{j=0}^{18} D_j \varpi^{2j+8},
\end{align}
with
\begin{align}\label{eq:detHfouDnull}
D_0 &=- 2194302107987911471104, \\ 
\label{eq:detHhourDeins}
D_1 & =- 5722539809029202405376,\\ 
\label{eq:detHfourDzwei}
D_2 &=- 6613012359516668960000,\\ 
\label{eq:detHfourDdrei}
D_3 & =- 4486664421458423037184, \\
\label{eq:detHfourDvier}
D_4 & =- 2001088031779516779472, \\
\label{eq:detHfourDfuenf}
D_5 & =- 623790394528514282984, \\
\label{eq:detHfourDsechs}
D_6 & =- 141205209279440754839, \\
\label{eq:detHfourDsieben}
D_7 & =- 23808533231074266904, \\
\label{eq:detHfourDvacht}
D_8 & =- 3041182357980945318, \\
\label{eq:detHfourDneun}
D_9 & =- 297427880736874480, \\
\label{eq:detHfourDzehn}
D_{10} & =- 22388520649376121, \\
\label{eq:detHfourDelf}
D_{11} & =- 1297313989039664, \\
\label{eq:detHfourDzwoelf}
D_{12} & =- 57565369848100, \\
\label{eq:detHfourDdreizehn}
D_{13} & =- 1933131508936, \\
\label{eq:detHfourDvierzehn}
D_{14} & =- 48125134553, \\
\label{eq:detHfourDfuenfzehn}
D_{15} & =- 858109496, \\
\label{eq:detHfourDsechzehn}
D_{16} & =- 10330246,  \\
\label{eq:detHfourDsiebzehn}
D_{17} & =- 74976, \\
\label{eq:detHfourDachtzehn}
D_{18} & =- 247. 
\end{align}

\end{appendices}


\newpage

\begin{thebibliography}{[99999]}
\footnotesize{
\vspace{-3.5pt}
  
\bibitem[AD]{AllenDynes} 
Allen, P. B., and Dynes, R. C.,
 \textsl{Transition temperature of strong-coupled superconductors reanalyzed},
 Phys. Rev. B \textbf{12}, 905--922 (1975).
\vspace{-3.5pt}
 
\bibitem[AM]{AllenMitrovic}
Allen, P. B., and Mitrovic, B., 
\textsl{Theory of superconducting $T_c$},
 Solid State Phys. \textbf{37}, 1--92 (1982).
\vspace{-3.5pt}
 
\bibitem[BR]{BergmannRainer}
Bergmann, G., and Rainer, D.,
 \textsl{The sensitivity of the transition temperature to changes in $\alpha^2\!F(\omega)$},
 Z. Phys. \textbf{263}, 59--68 (1973). 
\vspace{-3.5pt}
 
\bibitem[Ca]{Carbotte}
Carbotte, J. P.,
 \textsl{Properties of boson-exchange superconductors},
Rev. Mod. Phys. \textbf{62}, 1027--1157 (1990).
\vspace{-3.5pt}

\bibitem[CAEK]{CAEK}
Chubukov, A. V., Abanov, Ar. G., Esterlis, I., and Kivelson, S. A.,
\textsl{Eliashberg theory of phonon-mediated superconductivity -- When it is valid and how it breaks down},
 Ann. Phys. \textbf{417}, 168190 (2020).
\vspace{-15pt}

\bibitem[CAWW]{ChubukovETal}
Chubukov, A. V., Abanov, Ar. G., Wang,  Y.,  Wu, Y.-M.,
\textsl{The interplay between superconductivity and non-Fermi liquid at a quantum critical point in a metal}, 
 Ann. Phys. \textbf{417}, 168142 (2020).
\vspace{-15pt}

\bibitem[E]{Eliashberg} 
Eliashberg, G. M., 
\textsl{Interactions between Electrons and Lattice Vibrations in a Superconductor}, 
Zh. Eksp. Teor. Fiz. \textbf{38}, 966--976 (1960)  [Sov. Phys.--JETP \textbf{11}, 696--702 (1960)].
\vspace{-3.5pt}


\bibitem[Eetal]{EsterlisETal} 
Esterlis, I., Nosarzewski, B., Huang, E. W., Moritz, B., Devereaux, T. P., Scalapino, D. J., and Kivelson, S. A.,
\textsl{Breakdown of the Migdal--Eliashberg theory: a determinant quantum Monte Carlo study},
 Phys. Rev. B \textbf{97}, 140501 (2018).
\vspace{-3.5pt}

\bibitem[EKS]{EKS} 
  Esterlis, I., Kivelson, S. A., and Scalapino, D. J.,
\textsl{A bound on the superconducting transition temperature},
npj | Quant. Mater. \textbf{3}, 59 (2018).
\vspace{-3.5pt}

\bibitem[H1]{Holstein1}
Holstein, T., 
\textsl{Studies of polaron motion: Part I. The molecular-crystal model},
 Ann. Phys. \textbf{8}  325--342 (1959).
\vspace{-3.5pt}

\bibitem[H2]{Holstein2}
Holstein, T., 
\textsl{Studies of polaron motion: Part II. The ``small'' polaron}, 
Ann. Phys. \textbf{8} 343--389 (1959).
\vspace{-3.5pt}

\bibitem[K]{Kato}
Kato, T.,
 \textsl{Perturbation theory for linear operators},
 Springer, New York (1980).
\vspace{-3.5pt}
 
\bibitem[KAYa]{KAYgamma}
 Kiessling, M. K.-H., Altshuler, B. L., and  Yuzbashyan, E. A., 
 \textsl{Bounds on $T_c$ in the Eliashberg theory of superconductivity. I: The $\gamma$ model},
J. Statist. Phys. \textbf{192}, art.69, 35.pp (2025).
\vspace{-3.5pt}

\bibitem[KAYb]{KAYdispersive}
 Kiessling, M. K.-H., Altshuler, B. L., and  Yuzbashyan, E. A., 
 \textsl{Bounds on $T_c$ in the Eliashberg theory of superconductivity. II: Dispersive phonons},
44p., J. Statist. Phys. (in press, 2025).
\vspace{-3.5pt}

\bibitem[Ma]{Marsiglio} Marsiglio F., 
\textsl{Eliashberg theory: A short review}, 
 	Ann.  Phys. \textbf{417}, 168102 (2020).
\vspace{-3.5pt}

\bibitem[Mi]{migdal} Migdal, A. B., 
\textsl{Interaction between Electrons and Lattice Vibrations in a Normal Metal}, 
Zh. Eksp. Teor. Fiz. \textbf{34}, 1438--1446 (1958) [Sov. Phys.--JETP \textbf{7}, 996 -- 1001 (1958)].
\vspace{-3.5pt}

\bibitem[MC]{MoonChubukov} 
Moon, E.-G., Chubukov, A., 
\textsl{Quantum-critical Pairing with Varying Exponents}, J. Low Temp. Phys. \textbf{161}, 263--281 (2010).
\vspace{-3.5pt}

\bibitem[S]{Sad}  
Sadovskii, M. V.,
\textsl{Upper Limit for the Superconducting Transition Temperature in Eliashberg--McMillan Theory},
JETP Lett. \textbf{120}, 205--207 (2024).
\vspace{-3.45pt}


\bibitem[SAY]{Semenok} 
Semenok, D. V., Altshuler, B. L., Yuzbashyan, E. A., 
\textsl{Fundamental limits on the electron-phonon coupling and superconducting $T_c$}, 
arXiv:2407.12922 (2024). 
\vspace{-3.5pt}

\bibitem[Tra]{Tra} 	
Trachenko, K.,
Monserrat, B.,
Hutcheon, M., and
Pickard, C. J.,
\textsl{Upper bounds on the highest phonon frequency and superconducting temperature from fundamental physical constants}, 
arXiv:2406.08129 (2024).
\vspace{-3.45pt}


\bibitem[WAAYC]{WAAYC}
\!\!Wang, Y., Abanov, Ar. G.,  Altshuler, B. L.,  Yuzbashyan, E. A., and Chubukov, A. V.,
 \textsl{Superconductivity near a quantum-critical point: The special role of the first Matsubara frequency},
  Phys. Rev. Lett.  \textbf{117}, 157001 (2016).
\vspace{-3.5pt}
  
\bibitem[YAa]{YuzAltPRB2}
 Yuzbashyan, E. A., and Altshuler, B. L.,
 \textsl{Breakdown of the Migdal--Eliashberg theory and a theory of lattice-fermionic superfluidity},
 Phys. Rev. B \textbf{106}, 054518 (2022).
\vspace{-3.5pt}

\bibitem[YAb]{YuzAltPRB}
 Yuzbashyan, E. A., and Altshuler, B. L.,
 \textsl{Migdal--Eliashberg theory as a classical spin chain},
 Phys. Rev. B \textbf{106}, 014512 (2022).
\vspace{-15pt}

\bibitem[YAP]{YuzAltPat}
 Yuzbashyan, E. A., and Altshuler, B. L., and Patra, A.,
 \textsl{Instability of metals with respect to strong electron-phonon interaction}, 
arXiv:2409.19562 (2024).
\vspace{-3.5pt}

 \bibitem[YKA]{YuzKieAltPRB}
 Yuzbashyan, E. A., Kiessling, M. K.-H., and Altshuler, B. L.,
 \textsl{Superconductivity near a quantum critical point in the extreme retardation regime},
 Phys. Rev. B \textbf{106}, 064502 (2022).
}
\end{thebibliography}
\end{document}